\documentclass{aastex631}
\usepackage{amsmath,bm}

\begin{document}

\title{Vortex Dynamics in the Neutron Star Inner Crust}

\author[0009-0007-2375-2327]{Xin Sheng}
\affiliation{Physics Department, Columbia University, New York, NY 10027, USA}

\author[0000-0001-7108-3729]{Bennett Link}
\affiliation{Department of Physics, Montana State University, Bozeman, MT 59717, USA}

\author[0009-0006-2122-5606]{Matthew E. Caplan}
\affiliation{Department of Physics, Illinois State University, Normal, IL 61790, USA}
\affiliation{Department of Physics, University of Illinois Urbana–Champaign, Urbana, IL 61801, US} 

\author[0000-0002-6987-1299]{Yuri Levin}
\affiliation{Physics Department, Columbia University, New York, NY 10027, USA}



\begin{abstract}

We study the superfluid vortex motion in the neutron star inner crust through direct three-dimensional simulations of the coupled dynamics of the vortex and the nuclear lattice. We demonstrate the pinning of an initially moving vortex to the lattice through excitation of lattice vibrations, and show that the efficiency of this process is higher for attractive than for repulsive nucleus-vortex interactions. We explore the unpinning of a vortex under the action of the applied Magnus force, and find that it is influenced by multiple parameters, including the sign of the pinning force, the lattice orientation, composition, temperature, and the energy of pinning to individual nucleus. In lattices with multiple grains, the unpinning transition is triggered inside the grains with weaker pinning, propagates along the vortex (mediated by the excited Kelvin waves) and crosses into grains with stronger pinning. This is likely to effectively decrease the critical force at which vortices unpin and to produce extended regions of unpinned vorticity. Shearing of the crust lattice (e.g., by a starquake) initiates the unpinning of the vortices that are crossing the slip plane. A close encounter of an unpinned vortex with a pinned vortex would cause the latter to unpin, perhaps initiating 
an unpinning avalanche of many vortices.

\end{abstract}

\section{Introduction} \label{sec:intro}

Pulsar spin glitches are sudden increases in the rotation rate of a neutron star. To date, approximately 700 glitches have been observed \citep{Espinoza_2011,Basu_2021}, including large glitch events in the Vela pulsar \citep{McCulloch_1990} and PSR J0537–6910 \citep{Marshall_2004} with magnitude $\Delta \nu / \nu \sim 10^{-6}$. One glitch in the Vela pulsar took place in $<$12.5 s \citep{Ashton_2019}. 

\cite{Anderson_1975} suggested that spin glitches arise from a sudden coupling between the neutron star’s crust and the more rapidly rotating neutron superfluid that coexists with the crust. In this picture, the quantized vortices within the superfluid are strongly pinned to nuclei of the inner crust. As the crust spins down under electromagnetic torque, the superfluid retains its angular momentum to the extent that vortices remain pinned to the crust. As the difference in angular velocity between the superfluid and the crust increases, a Magnus force develops on the pinned vortices. When the Magnus force locally reaches a critical value, the vortices are forced to unpin and become mobile. If many vortices become unstable at once, their dissipative motion transfers angular momentum from the superfluid to the crust, driving a spin glitch.

Several aspects of the model of \cite{Anderson_1975} remain poorly understood. Firstly, reliable calculations of the strength to which vortices pin to the nuclear lattice have been elusive. The strength and sign (repulsive vs. attractive) of the vortex–lattice interaction has been tackled by several approaches, with significantly different results. \cite{Avogadro_2007} solved the mean-field Hartree–Fock–Bogoliubov (HFB) equations and reported pinning energies in the range $\sim -1$ to $4$ MeV depending on density. Subsequent work by the same group yielded pinning energies $\sim -4$ to $18$ MeV for SLy4 and SkM models in density range $\rho \sim 10^{12} - 10^{14}$ g/cm$^3$ \citep{Avogadro_2008}. \cite{Donati_2003} computed both the nuclear pinning and the interstitial pinning and reported $\sim -4 - 14$ MeV. In a different approach, \cite{Bolgac_2013} performed time-dependent Density Functional Theory (DFT) calculation, and obtained a bell-shaped relation between the pinning force and the distance $r$ between the vortex and the nucleus, with absolute maximum pinning force per unit length $\sim 0.08$ MeV/fm$^2$ at $r \sim 8$ fm with width $\approx 6$ fm. \cite{Wlazłowski_2016} also performed simulations with the DFT, but found the maximum pinning force per unit length $\sim -0.13$ MeV/fm$^2$ and decrease over distance for $r > 8$ fm. At this point it is clear that the pinning interaction is of order $\sim 1$ MeV over a length scale of $\sim 10$ fm, but no consensus has emerged on the density dependence or sign of the pinning interaction, nor the physical extent of the pinning region. 

Secondly, the dynamics of vortex pinning and unpinning has not been fully understood. Recent progress has been made on vortex motion in a static nuclear lattice in three dimensions \citep{Link_2009, Link_2022}, but how vortices move in a dynamic lattice requires further study. As vortices move through the nuclear lattice, lattice motion is excited, affecting the dynamics of pinning and unpinning, and producing drag between the superfluid and the nuclear lattice via the vortices. The drag force determines the rate at which the crust will be spun up by unpinned vortices \citep{Baym_1992, Jones_1992}. In this paper we address this problem with 3D simulations of the vortex motion coupled to the dynamic lattice.

Finally, the arguably most important question is how spin glitches are triggered in the first place. To drive a typical spin glitch in the Vela pulsar, $\sim 10^{11}$ vortices must move a sizable fraction of the crust thickness in under $\sim 10$ s. Ideas for the glitch trigger include direct unpinning of a many vortices by a starquake \citep{Ruderman_1991}, rapid thermal creep due to quake-induced crust heating \citep{Link_1996},\footnote{There is some tantalizing evidence for a crust quake trigger of the 2016 Vela glitch \citep{Palfreyman2018, Bransgrove_2020}, but more observations are needed to confirm this.} and vortex ``avalanches'', similar to those observed in laboratory superconductors \citep{Field_1995, Miclea_2009}.  \cite{Cheng_1988} proposed that as a single unpinned vortex approaches a pinned one, the latter could be forced to unpin, creating a chain reaction, and a vortex avalanche develops. This idea has been further explored with 2D simulations \citep{Warszawski_2012, Howitt_2020,liu2025vortex}. These simulations have two significant limitations. First, they fail to capture the rich dynamics of real vortices, which are extended objects in three dimensions. Second, these simulations are limited, for computational feasibility, to a dense vortex lattice regime in which the vortex separation is not much larger than the spacing of the nuclear lattice. In a neutron star, however, the mean vortex spacing is some 10 orders of magnitude larger than the spacing of the nuclear lattice, so the vortex lattice is effectively quite dilute. In this regime, vortices can only interact and trigger unpinning when they actually intersect or come close to one another. A detailed three-dimensional simulation of interacting vortex pairs is necessary in order to resolve the unpinning dynamics.

In this paper, we address some of the issues outlined above through direct numerical simulations. We build on the work of \cite{Link_2009} and especially \cite{Link_2022}, who developed a spectral code to study the three-dimensional vortex motion inside a rigidly fixed crustal lattice. To account for motion of the nuclear lattice, we carry out three‐dimensional molecular dynamics (MD) simulations of the vortex–lattice system. This approach allows us to resolve the lattice’s response and directly assess its influence on vortex pinning and unpinning. In \autoref{sec:method} we introduce our simulation setup and explain our numerical techniques. The results of our simulations are presented in \autoref{sec:result}, where we study the pinning and unpinning of a single vortex, as well as the interaction between the vortices. In \autoref{sec:discussion}, we discuss the implications of our
results.

\section{Method}
\label{sec:method}

To simulate the vortices in the neutron star crust, we couple two pieces of physics: the motion of a neutron star vortex under an external force density distributed along the vortex, and the motion of the lattice. We describe the methods in turn below. The unit system used in this paper is described in \autoref{tab:simulation_unit}.

\subsection{Vortex equation of motion}

We treat vortices in classical continuum dynamics. We assume that the vortices are approximately aligned with an arbitrary $z$ axis, and specify the shape of the vortex by a two-dimensional displacement vector $\bm{u}(z, t) = u_x \hat{x} + u_y \hat{y}$. For small bending angles, $|\partial \bm{u}/\partial z| \ll 1$, the bending force is proportional to $\partial^2 \bm{u}/\partial z^2$. We adopt the equation of motion from \cite{Link_2022} and importantly, remove their artificial damping term to obtain 
\begin{equation}
\label{eqn:eom}
    T_v \frac{\partial^2 \bm{u}}{\partial z^2} + \rho_s \bm{\kappa} \times \left(\frac{\partial \bm{u}}{\partial t} - \bm{v_b}\right) + \bm{f_{\mathrm{L}}} = 0,
\end{equation}
where $T_v$ is the vortex tension, $\rho_s$ is the superfluid mass density, $\bm{\kappa}$ is the local vorticity vector with magnitude $|\bm{\kappa}| = h/2m_n$, $\bm{v_v} \equiv \partial \bm{u} /\partial t$ is the local vortex velocity, $\bm{f_\mathrm{L}}$ is the external force from the lattice, and $\bm{v_b}$ is the background superfluid velocity. Followed \cite{Link_2022}, we adopt $\rho_s=10^{13}$ g/cm$^{3}$ and $T_v = 0.6$ MeV/fm in this paper. The first term in \autoref{eqn:eom} is the bending force of the vortex. The second term is the Magnus force computed through the velocity difference of vortex itself and its surrounding velocity field. The last term is the external force from the lattice. For multi-vortex system, the superfluid velocity on one vortex will be determined by a superposition of that generated by nearby vortices, and the background superfluid velocity. The equation of motion is solved using spectral methods described in \autoref{Appendix:solver}. 
\subsection{Lattice setup}

The neutron star inner crust is treated as a body-centered-cubic (bcc) Coulomb lattice \citep{Chamel_2008}. The lattice is simulated by \texttt{LAMMPS}, an effective MD simulation tool for multi-particle systems \citep{LAMMPS}. The equations of motion of the lattice are given in Section 2 of \cite{Shinoda_2004}. A typical choice of the lattice spacing is $b=30$ fm \citep{Negele_1973} \footnote{The `lattice spacing' in this paper refers to the cubic box side length for the bcc unit cell. The Wigner–Seitz radius for this lattice is $a_i = 14.77$ fm.} . The interaction potential between nuclei is given by the Coulomb force with ultra-relativistic Thomas-Fermi screening by electrons: $V_{\mathrm{lattice}} = C(Ze)^2\exp(-r/\lambda)/r$, where $Z$ is the nuclear charge and $\lambda$ is the screening length. In our simulation, $\lambda = 23.3$ fm for $Z=40$. 

We introduce a Nosé–Hoover thermostat \citep{Nose_1984, Hoover_1985} to control the temperature of the system. We choose the timescale on which the thermostat is coupled to the lattice to be comparable to the thermal diffusion timescale $t_{\mathrm{diff}} \sim L_b^2 C_V/\kappa_T\sim10^{-19}$ s, where $L_b$ is the length-scale of the simulation box, $C_V$ is the specific heat, and $\kappa_T$ is the thermal conductivity \citep{Gnedin_2001}. We have tested that our results are insensitive to this choice.

The time integration is performed by Stoermer-Verlet algorithm, and the timestep is adaptive and controlled by the vortex integration (see \autoref{Appendix:solver}). Our \texttt{LAMMPS} approach is derived from recent MD work in \cite{2024PhRvL.133m5301C,2025RNAAS...9..162C}. We use a variety of box sizes, ranging from $20 \times 20 \times 40$ to $40 \times 40 \times 40$. We numerically solve for the vortex configuration $\bm{u}(z,t)$, nuclear positions $\bm{R_k}(t)$, the force exerted by the vortex on each nucleus, $\bm{f_{\mathrm{vk}}}(\bm{R_k}, \bm{u}, t)$, and the force per unit length exerted by the nuclei on the vortex, $\bm{f_{\mathrm{L}}}(z,t)$.

\subsection{Vortex-Lattice interaction}

Following \cite{Link_2022}, we assume that each vortex element at location $\bm{u}(z)$ feels a Gaussian potential from each nucleus,
\begin{equation}
    V(z) = \sum_k V_k(\bm{R_k}, z) = \sum_k \frac{E_i}{b} \exp \left[- \left(\frac{|\bm{R_{k}} - \bm{u}(z)|}{\sigma}\right)^2\right]
\end{equation}
where $E_i$ is the interaction energy, $b$ is the lattice spacing, and $\sigma$ is the length scale of the interaction. The interaction length-scale $\sigma$ is comparable to the vortex core radius, which we fix at 6 fm=0.2$b$. The force per unit length on the vortex is given by $\bm{f_\mathrm{L}}(z) = -\nabla_{\perp}V(z)$, where $\nabla_{\perp}$ is the gradient perpendicular to the vortex. For a single nucleus $k$, the total force exerted by the vortex on that nucleus $\bm{f_{\mathrm{vk}}}(\bm{R_k})$ is obtained by integrating along the vortex,
\begin{equation}
    \bm{f_{\mathrm{vk}}}(\bm{R_k}) = \int_{\bm{u}} \nabla_{\perp}V_k(\bm{R_k},z) \, dz.
\end{equation}
The corresponding pinning energy $E_p$ is the work required to move the vortex from $r=0$ to $r=\infty$ against the 
radial component of this force, $f_{\mathrm{vk}}(r)$:
\begin{equation}
    E_p = \int_0^{\infty} f_{\mathrm{vk}}(r) dr = \frac{\sqrt{\pi}\sigma E_i}{b},
\end{equation}
giving $E_p \approx 0.35E_i$ in our simulation. 


\begin{table}[]
    \centering
    \begin{tabular}{c|c|c}
     Physical variables   & Simulation Unit & Physical unit\\
     Length &     $b^*$    & $30$ fm \\
     Time    & $t^* = \rho_s \kappa b^{*2}/T_v$ & $1.86 \times 10^{-20}$ s \\
     Velocity & $v^* = b^*/t^*$ & $1.61 \times 10^{8}$ cm/s \\
    \end{tabular}
    \caption{The simulation units and corresponding physical units used in this paper.}
    \label{tab:simulation_unit}
\end{table}

\section{Simulations}
\label{sec:result}

\subsection{Vortex Relaxation and pinning without background flow}

\begin{figure}
    \centering
    \includegraphics[width=0.9\linewidth]{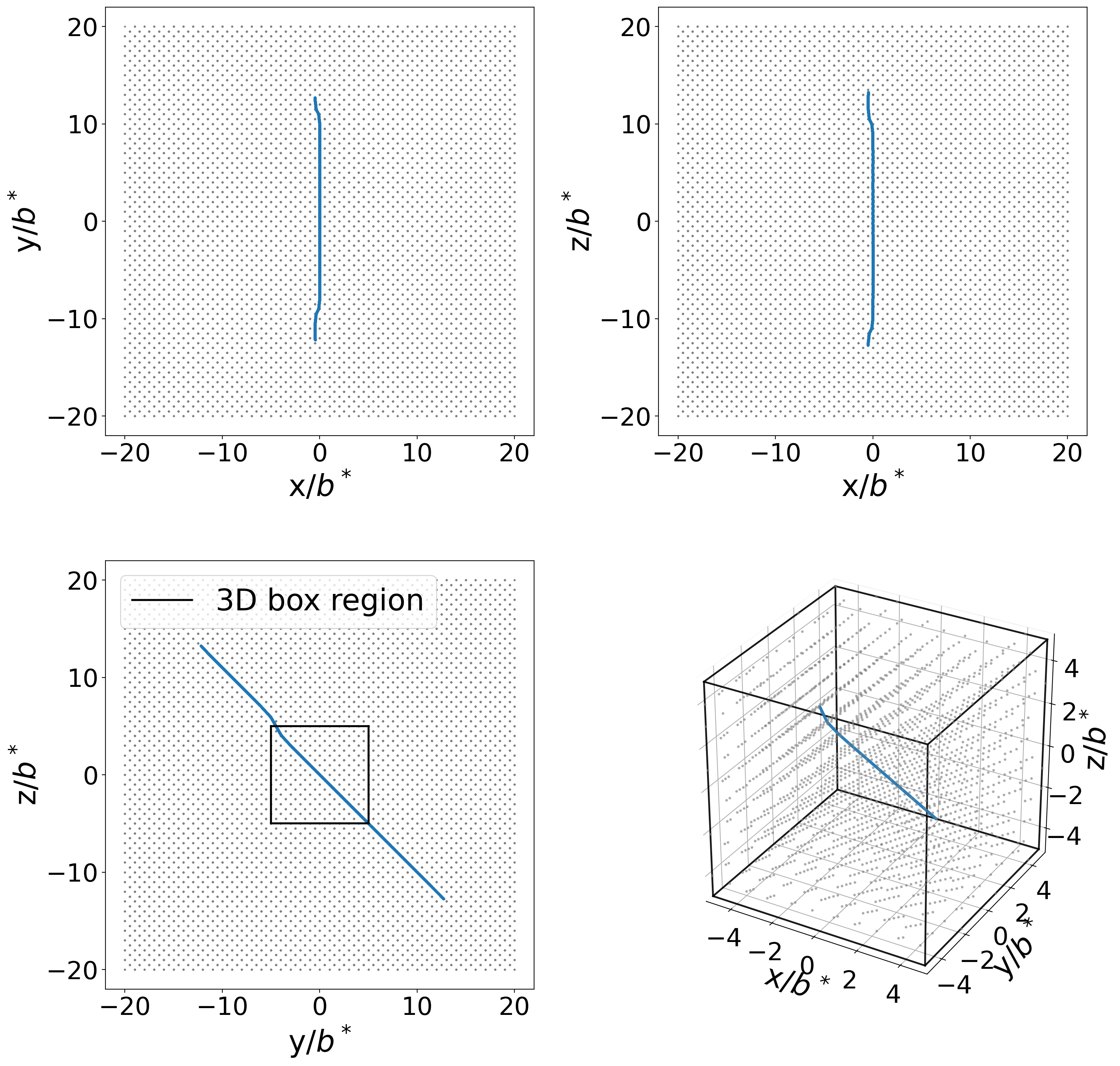}
    \caption{One sample snapshot in the relaxation simulation to illustrate the 3-D vortex-lattice system. We show different views of the box, including the $xy$-plane (Top-left), $xz$-plane (Top-right), and the $yz$-plane (Bottom-left). We also show a zoom-in 3D box in the simulation region (Bottom-right). The line is the vortex, and the dots are nuclei. }
    \label{fig:sample}
\end{figure}

\begin{figure}
    \centering
    \includegraphics[width=0.48\linewidth]{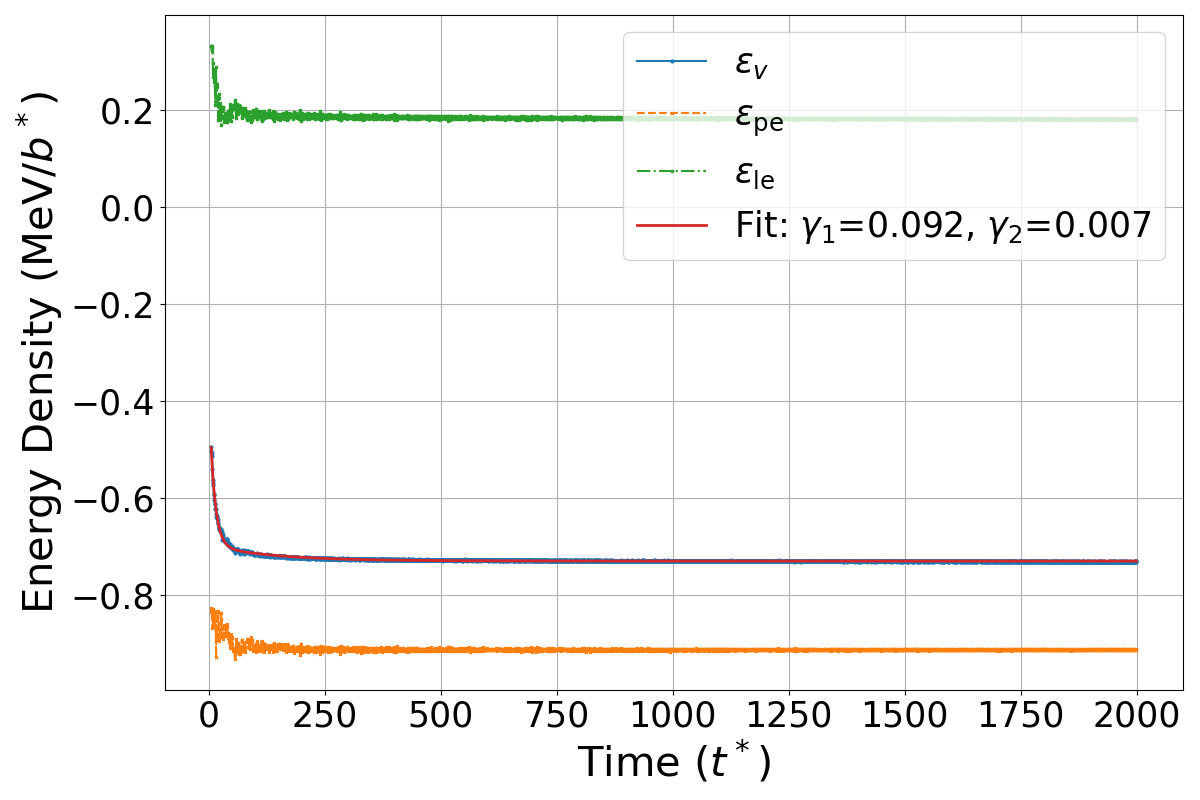}
    \includegraphics[width=0.48\linewidth]{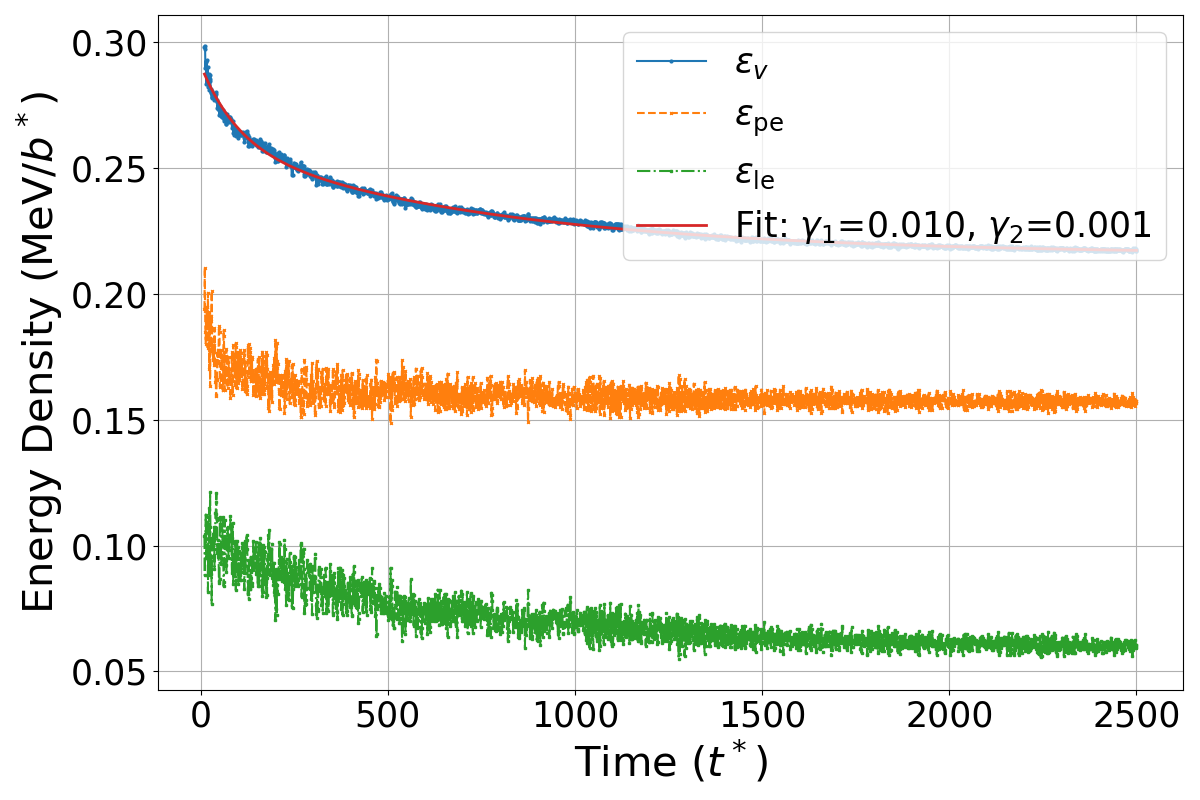}
    \caption{The line energy density $\epsilon_{\mathrm{le}}$, potential energy density $\epsilon_{\mathrm{pe}}$ and total energy density $\epsilon_{v}$ of the vortex vs. time in relaxation for attractive (left) and repulsive (right) interaction. The vortex undergo an energy dissipation and reach equilibrium gradually. The vortex relaxes much faster in attractive interaction compared to repulsive interaction. We fit the energy damping with a bi-exponential model $E_v = E_1e^{-\gamma_1t} + E_2e^{-\gamma_2t} + C$. We find out that on average $\gamma_1\sim 0.1$ for attractive interaction and $\gamma_1\sim0.01$ for repulsive interaction. }
    \label{fig:relaxation}
\end{figure}

We begin by investigating how an initially unpinned vortex gets pinned to the lattice. In a realistic lattice, nuclei move under the forces exerted on them by the vortex, dissipating energy of the Kelvin waves and ultimately leading the vortex to come to rest in a mechanical equilibrium. In our simulations, the slow-down of the vortex motion and its pinning to the lattice is enabled by the energy exchange between the vortex and the lattice. In our initial configuration, the vortex features a multitude of Kelvin waves which are propagating along its extent. We examine the vortex’s relaxation behavior for its various orientation directions relative to the lattice, and for random initial conditions. The are two contributions to the vortex length: 1) the interaction energy with the lattice, and, 2) the bending energy:

\begin{equation}
    E_v = E_{\mathrm{pe}} + E_{\mathrm{le}} =\int_L \sum_i \frac{E_i}{b} \exp \left[- \left(\frac{|\bm{R}_i-\bm{u}(z)|}{\sigma}\right)^2\right] dz + \frac{T_v}{2} \int_L \left(\frac{\partial \bm{u}(z)}{\partial z}\right)^2 dz.
\end{equation}
where $\bm{R_i}$ is the position of nucleus $i$. The second term is the energy per unit length of the vortex $T_v$ times the increase in length with respect to a straight vortex. As the vortex damps to a pinned configuration, $E_v$ reaches a local minimum. The rate at which a vortex pins is determined by the rate at which the energy is transferred from the vortex to the lattice. To estimate this rate, we randomly select ten different vortex orientation directions, with random perturbations around the straight-line shape. We consider both attractive and repulsive vortex-lattice forces. A sample snapshot from the simulation is shown in \autoref{fig:sample}, and two of the example results are shown in \autoref{fig:relaxation}. The relaxation proceeds in two stages. (i) At early times the range of its motion exceeds the typical lattice spacing and the vortex loses energy rapidly due to the strong vortex-nucleus interaction. (ii) After the initial relaxation, pins to some number of nuclei, but continues small amplitude motion which decays on a longer timescale. We capture this behavior with a bi-exponential model,
\begin{equation}
    E_v \approx E_1e^{-\gamma_1t} + E_2e^{-\gamma_2t} + C
\end{equation}
where $\gamma_1$ describes the fast damping and $\gamma_2$ describes the slow equilibration. In practice, a single-exponential decay of the first stage is sufficient to describe the pinning process. We find that when the vortex-lattice force is attractive, the vortex in its pinned  configuration passes close to some nuclei, and the energy exchange between the vortex and the lattice is efficient. By contrast, when the vortex-lattice force is repulsive, the vortex is repelled from the nucleus and pins to the interstices of the lattice \citep{Link_1991}. As a result, the effective damping rate is smaller than that for the attractive force. We find that $\gamma_1 \sim 0.1$ for attractive interaction and $\gamma_1 \sim 0.01$ for repulsive interaction.

\subsection{Single Vortex unpinning and repinning}

Our study focuses on determining the critical unpinning velocity and understanding the repinning dynamics accounting for lattice dynamics. Given that the typical spacing between vortices ($10^{-4}$ m) is much larger than the length scale of the self-induced velocity field ($10^2$ fm), we can treat each vortex independently, provided that their trajectories do not intersect. We investigate both unpinning and repinning behavior under attractive and repulsive interactions, considering a range of influencing factors. We begin by considering the dynamics of an isolated vortex, moving through a dynamic lattice.

\subsubsection{Baseline model}

\begin{figure}
    \centering
    \includegraphics[width=0.98\linewidth]{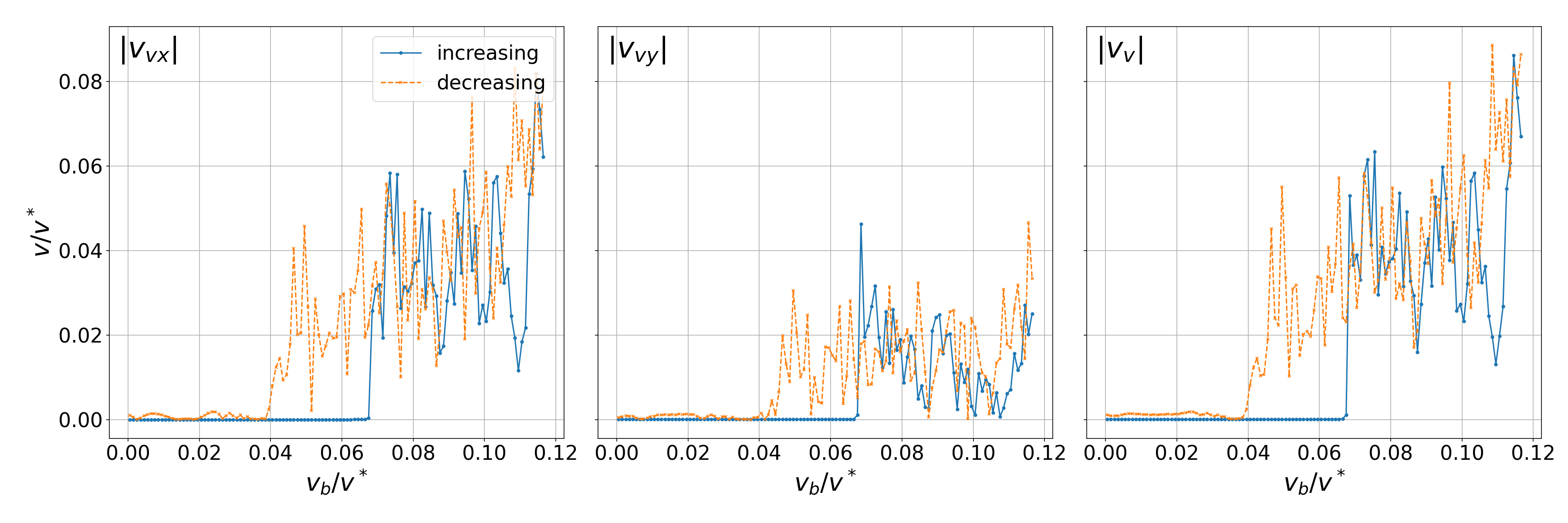}
    \includegraphics[width=0.98\linewidth]{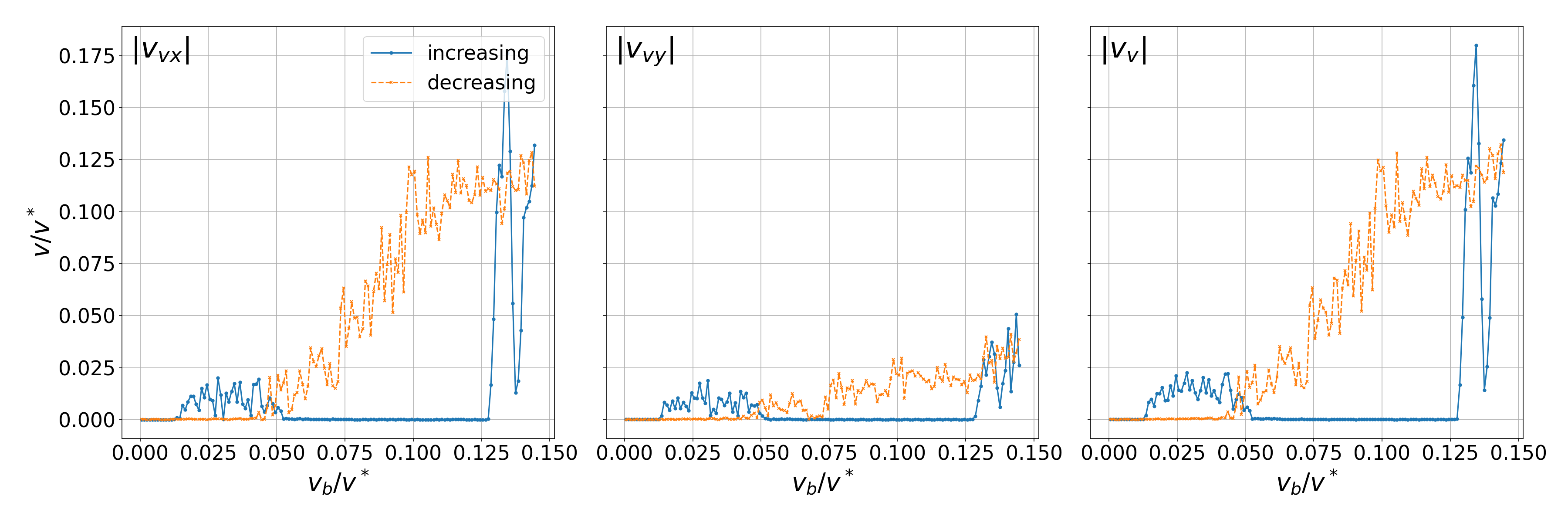}
    \caption{Two selected baseline model simulations showing the vortex velocity in x direction $|v_{vx}|$, in y direction $|v_{vy}|$, and absolute value $|v_v|$ vs. the background superfluid velocity $v_b$. The solid lines show the vortex velocity when increasing $v_b$, and the dashed lines show the vortex velocity when decreasing $v_b$. These two models have different orientations, which can affect the unpinning velocity significantly. The vortex velocity shows hysteresis, where the superfluid velocity at unpinning is larger than that at repinning. The results are qualitatively similar to those in \cite{Link_2022}.}
    \label{fig:baseline}
\end{figure}

To set a baseline model for pinning and unpinning, we adopt the following setup in the simulation with attractive interaction: the lattice has nuclear charge $Z=40$ and atomic nucleus mass $M=200$, with lattice spacing $b=30$ fm and interaction energy $E_i = 4$ MeV (corresponding to $E_p = 1.4$ MeV). We start at a low temperature $T = 3 \times 10^3$ K. 
We launch ten simulations with randomly selected lattice orientations. First, the Magnus force is set to zero and the vortex starts from a straight initial position and then undergoes relaxation from $t=0$ to $t=400t^*$ with $v_b = 0$. To accelerate the relaxation, we introduce an artificial damping $\gamma = 0.1$ in the first 200 timesteps, which is removed before the relaxation ends. After relaxation, we linearly increase the $v_b$ until the vortex unpins and moves a significant distance (all $v_b$ is on $+x$ direction in the frame of the vortex, $\bm{v_b} = v_{b}\hat{x}$). We then gradually decrease $v_b$ to zero. Two of the examples are shown in \autoref{fig:baseline}. The axes are chosen so that $v_{vx}$ is in the direction of $v_b$, while $v_{vy}$ is perpendicular to $v_b$. In some cases, the vortex re-adjusts itself to a stronger pinning position before it fully unpins. After the vortex gets unpinned, $v_{vx}$ increases linearly with $v_b$ while $v_{vy}$ keeps relatively constant until the vortex get repinned. The unpinning velocity range is $v_c \sim 10^6 -10^7$ cm/s.

Similar to \cite{Link_2022}, we find strong hysteresis behavior: the vortex will not repin  until $v_b$ decreases to a small value compared to the unpinning velocity $v_c$. The critical velocity for repinning $v_r$ is smaller than $v_c$. The reason is that the unpinned vortex has a high degree of Kelvin-wave excitations that inhibits the vortex from re-pinning. 

\begin{figure}
    \centering
    \includegraphics[width=0.45\linewidth]{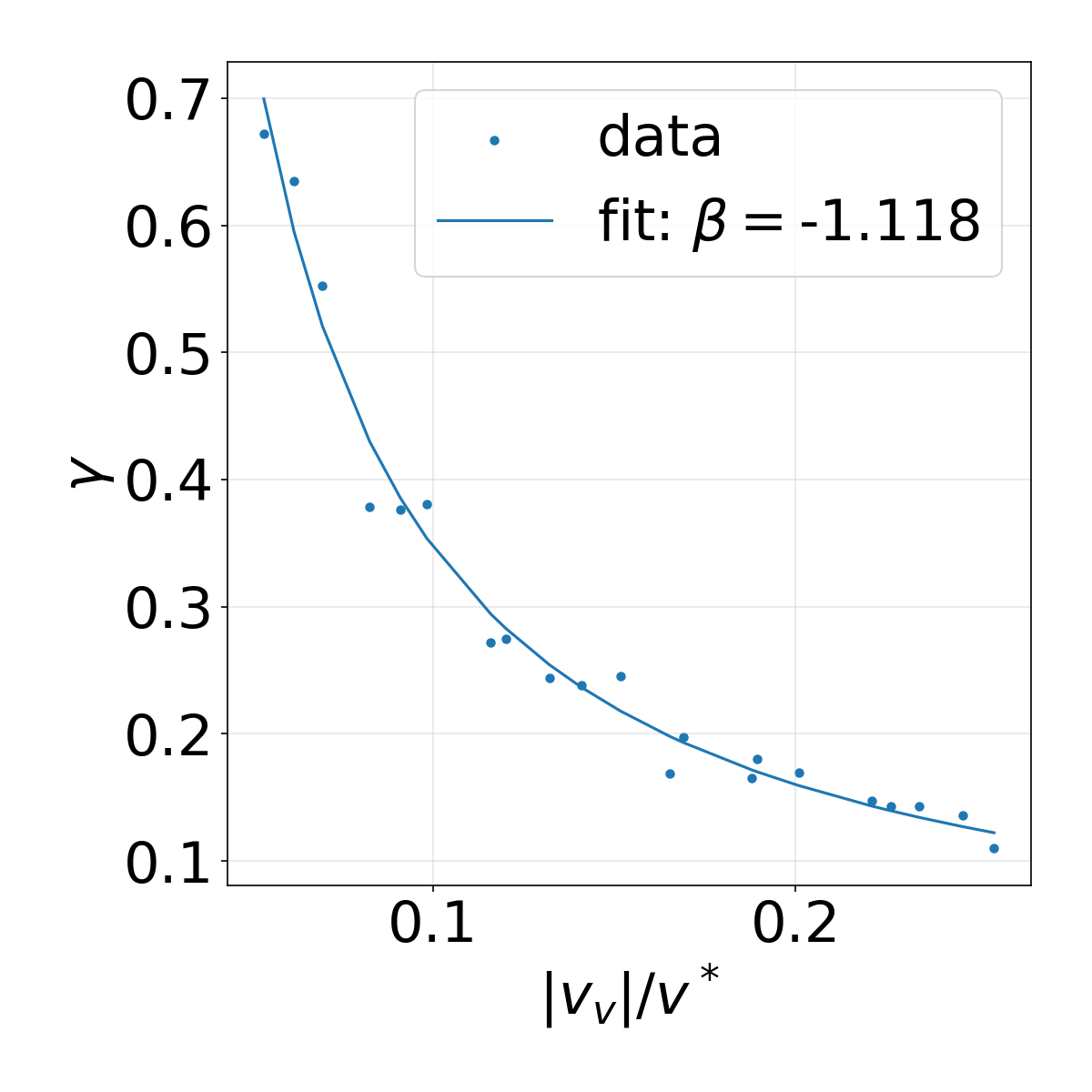}
    \includegraphics[width=0.45\linewidth]{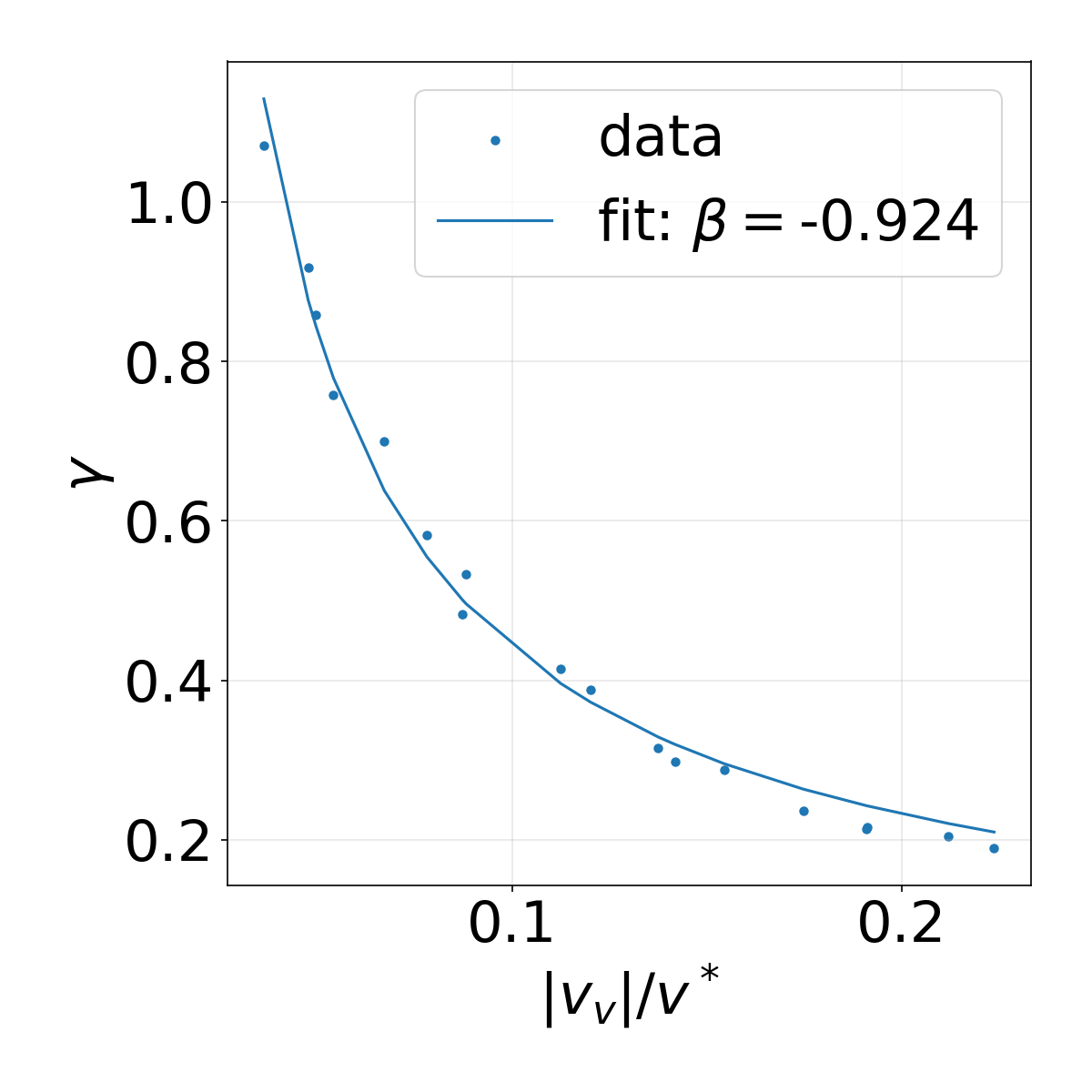}
    \caption{Damping coefficient $\gamma$ vs. vortex  velocity $v_v$ for two selected simulations in attractive interaction. We assume $\gamma \propto v_v^\beta$ and find $\beta \sim -1$. }
    \label{fig:gamma}
\end{figure}

To understand the long-term dynamics of the vortex, we evaluate the effective drag force acting on it as a result of its dissipative interaction with lattice. As an unpinned vortex moves through the lattice, the bending term in \autoref{eqn:eom} averages to zero, and the average velocity $\bm{v_v}$ of the vortex is given by 
\begin{equation}
    \rho_s\bm{\kappa}\times(\bm{v_v}-\bm{v_b})-\eta\bm{v_v}=0
\end{equation}

\noindent where we define $\gamma \equiv \eta/\rho_s \kappa$. The drag term accounts for the transfer of energy from the vortex to lattice vibrations. The dissipated power per unit vortex length is given by $P = \eta v_v^2 = E_s v_v/b$, where $E_s$ is the average scattering energy per nucleus \citep{Baym_1992}. These authors parameterized $\gamma \propto E_s/v_v \propto v_v^{\beta}$ and analytically found $\beta = -3/2$ by assuming weak pinning. The calculation was in a high $v_v$ limit; there is no pinning and that the energy transfer between the lattice and vortex is small. Our simulations take place in a different regime that cannot be treated analytically, and we do not find this scaling. \autoref{fig:gamma} shows our measurement of $\gamma$ for two simulations. We plot $\gamma$ vs. $v_v$, fit the curve by assuming $\gamma \propto v_v^\beta$. We find $\beta \sim -1$.

\subsubsection{Crust Model}
\label{subsubsec:crust_model}

In the neutron star inner crust, the nuclear charge $Z$ sets the Coulomb interaction strength between nuclei, while the nuclear mass number A governs each nucleus’s inertial response. In ground‐state crust models, $Z$ typically lies between 30 and 40 \citep{Chamel_2008}, whereas in accreted‐crust scenarios $Z$ can be as low as 10 \citep{Haensel_2003}. We examine four representative models for compositions:

\begin{enumerate}
\item Accreted‐crust model: $Z = 10$, $A = 40$, $\lambda = 37.0$ fm.
\item Light‐mass ground‐state: $Z = 30$, $A = 100$, $\lambda = 25.6$ fm.
\item Mid‐mass ground‐state: $Z = 35$, $A = 150$, $\lambda = 24.4$ fm.
\item Heavy‐mass ground‐state: $Z = 40$, $A = 200$, $\lambda = 23.3$ fm.
\end{enumerate}

\begin{figure}
    \centering
    \includegraphics[width=0.45\linewidth]{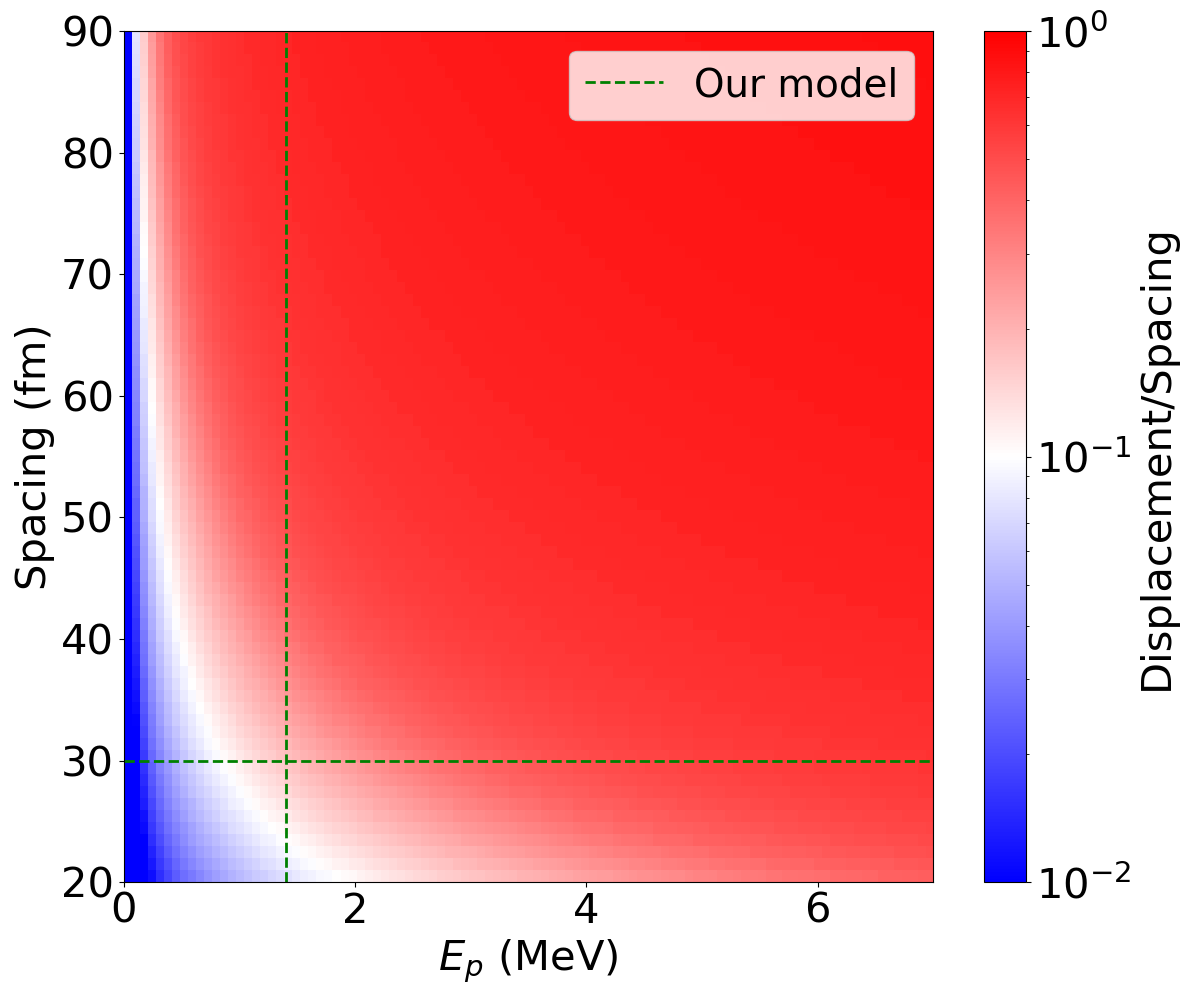}
    \includegraphics[width=0.45\linewidth]{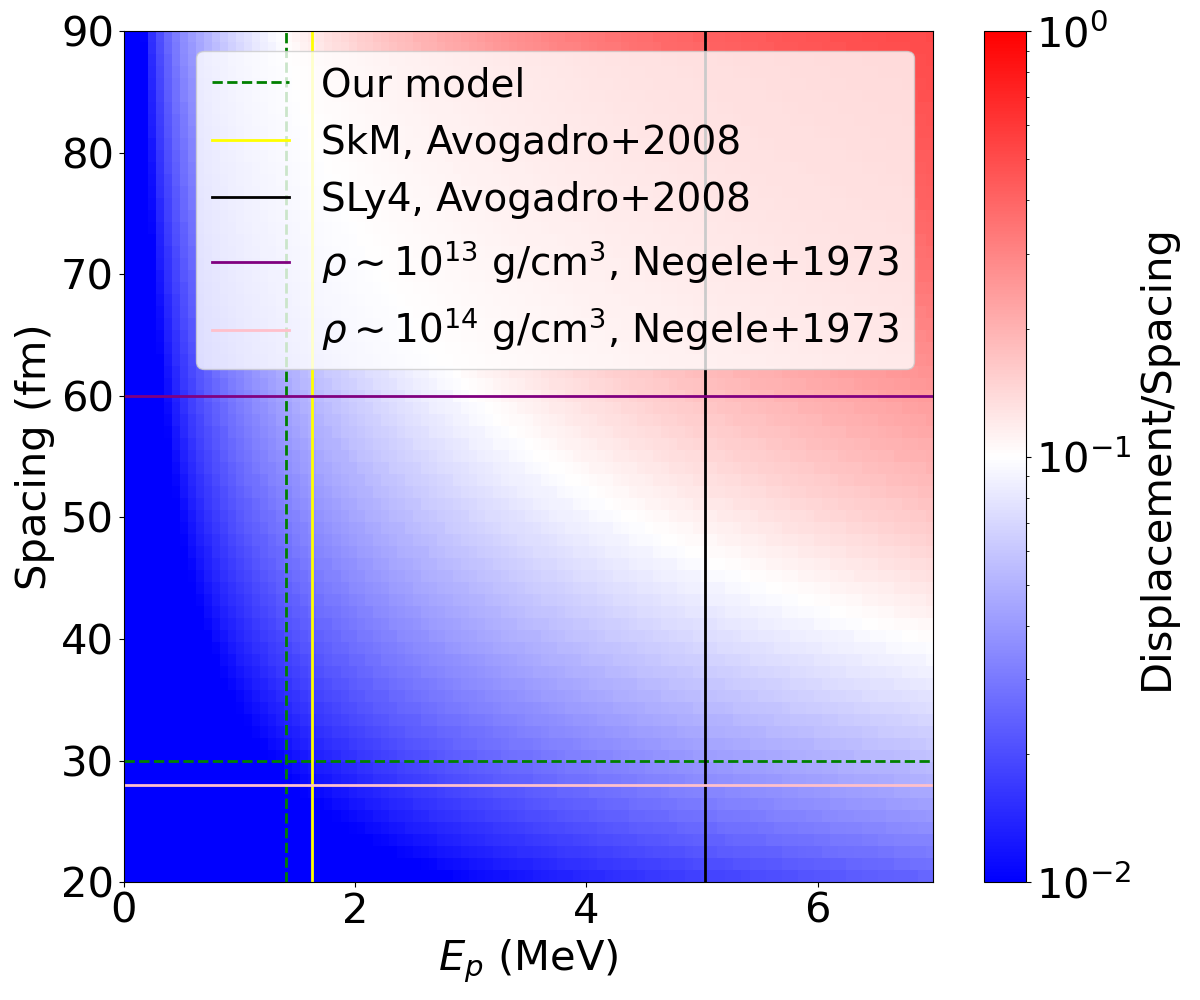}
    \caption{Lattice displacement in different lattice spacing and vortex pinning energy $E_p$. The displacement is measured by finding the balance between the pinning force and Coulomb force. Left: lattice with nuclear charge $Z=10$; Right: $Z=40$. Dashed lines indicate our choice of the baseline model. Two horizontals solid lines show the lattice spacing with different nucleus density computed in \cite{Negele_1973}. Two vertical solid lines show the $|E_p|$ for two models computed in \cite{Avogadro_2008}. A displacement $> 0.1$ indicates possible non-harmonic deformation and restoration. If the Coulomb force is weak compared to the vortex interaction, we observe large lattice deformation where a fully coupled MD simulation is required; otherwise, the lattice can be well-approximated as stationary. }
    \label{fig:displacement}
\end{figure}

A vortex deforms the nuclear lattice. We estimate the characteristic deformation as follows. We consider a regular lattice with all but one of the sites fixed in space. We then apply the maximum force from the vortex-nucleus interaction to the deformable lattice, and compute the displacement. The result is shown in \autoref{fig:displacement}. The estimate indicates that the lattice remain essentially rigid across all ground‐state compositions at $b = 30 \, \mathrm{fm}$, whereas the accreted‐crust model exhibits appreciable deformation. 

\begin{figure}
    \centering
    \includegraphics[width=0.5\linewidth]{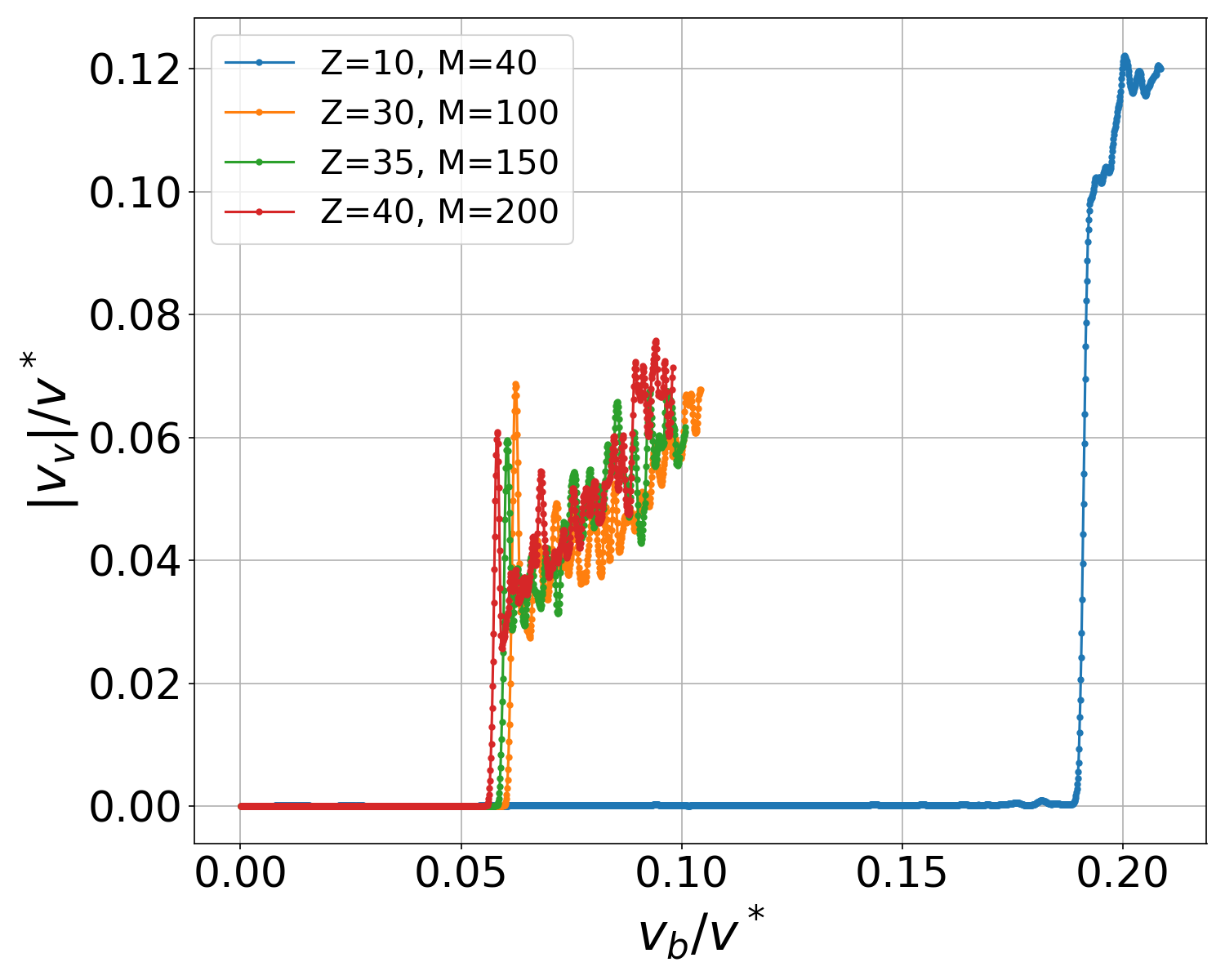}
    \caption{Vortex velocity $|v_v|$ vs. background superfluid velocity $v_b$ showing the pinning transition of the vortex for different nuclear charge $Z$ and nucleus mass $A$ with $b=30$ fm. For accretion model where $Z \sim 10$, the lattice can be largely deformed (see \autoref{fig:displacement}) and unpinning velocity can be significantly larger, whereas for ground-state model, the unpinning velocity is not very sensitive to the nuclear charge in range $Z \sim 30 - 40$. }
    \label{fig:Z=10}
\end{figure}

Our numerical simulations confirm that the critical unpinning velocity varies only marginally among the ground‐state cases $2 - 4$ with the same lattice spacing. In the accreted-crust model shown in \autoref{fig:Z=10}, the unpinning velocity shows a significant increase, as a vortex pulls nuclei toward it, enhancing the strength of pinning.

\subsubsection{Temperature}

\begin{figure}
    \centering
    \includegraphics[width=0.5\linewidth]{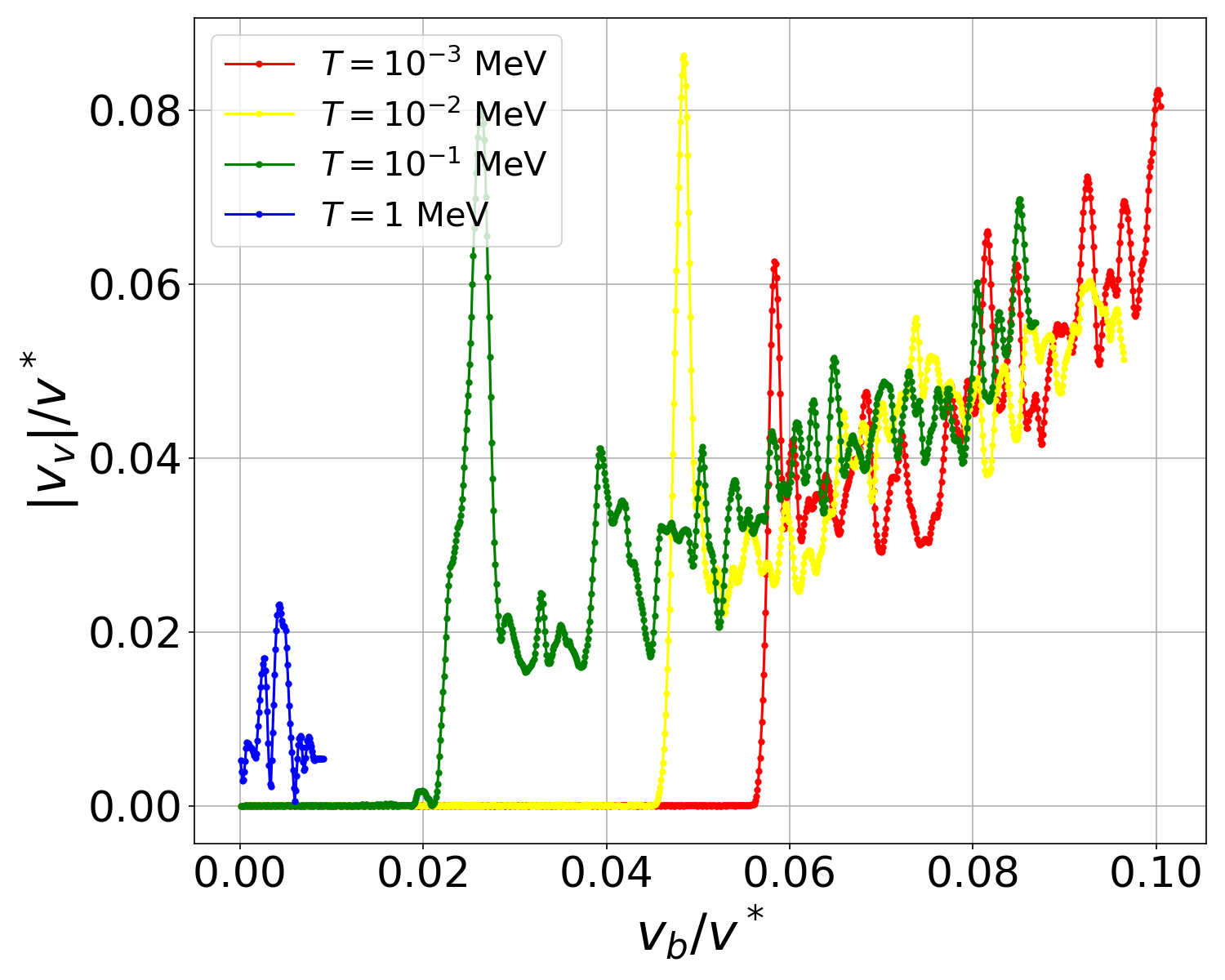}
    \caption{Vortex velocity $|v_v|$ vs. background superfluid velocity $v_b$ showing the pinning transition of the vortex for $T=10^{-3}-1$ MeV ($10^7 - 10^{10}$ K). Below $10^{-3}$ MeV (not shown in the plot) the temperature will not affect the unpinning velocity and the dynamics is similar to the $10^{-3}$ MeV case. Higher temperature will excite the lattice oscillation and effectively reduce the pinning strength. }
    \label{fig:T=1e9}
\end{figure}

A neutron star cools after its birth in the supernova explosion. The crust forms at temperature around 1 MeV ($10^{10}$ K). To  explore thermal effects on pinning pinning dynamics, we perform simulations over a temperature range $10^{-3}$ MeV ($10^7$ K) to 1 MeV ($10^{10}$ K). All of the simulations have the same initial conditions except the temperature. We show an example in \autoref{fig:T=1e9}. The unpinning velocity for this example is $v_c \sim 10^7$ cm/s ($v_c/v^* \sim 0.06$) until the temperature reaches 0.01 MeV ($10^8$ K), where the lattice oscillation becomes significant. Above 0.01 MeV, the unpinning velocity drops and the pinning becomes weaker. For $T = 1$ MeV, the critical superfluid velocity for unpinning $v_c$ drops significantly and thermally-excited lattice motion strongly influences the unpinning dynamics. Near the melting temperature $T_m \sim 1$ MeV the lattice thermal oscillations of the lattice becomes so large that pinning effectively vanishes. Our simulation shows that the pinning can be significantly influenced by the temperature when $T> 0.01$ MeV. 

\subsubsection{Pinning energy}

\begin{figure}
    \centering
    \includegraphics[width=0.5\linewidth]{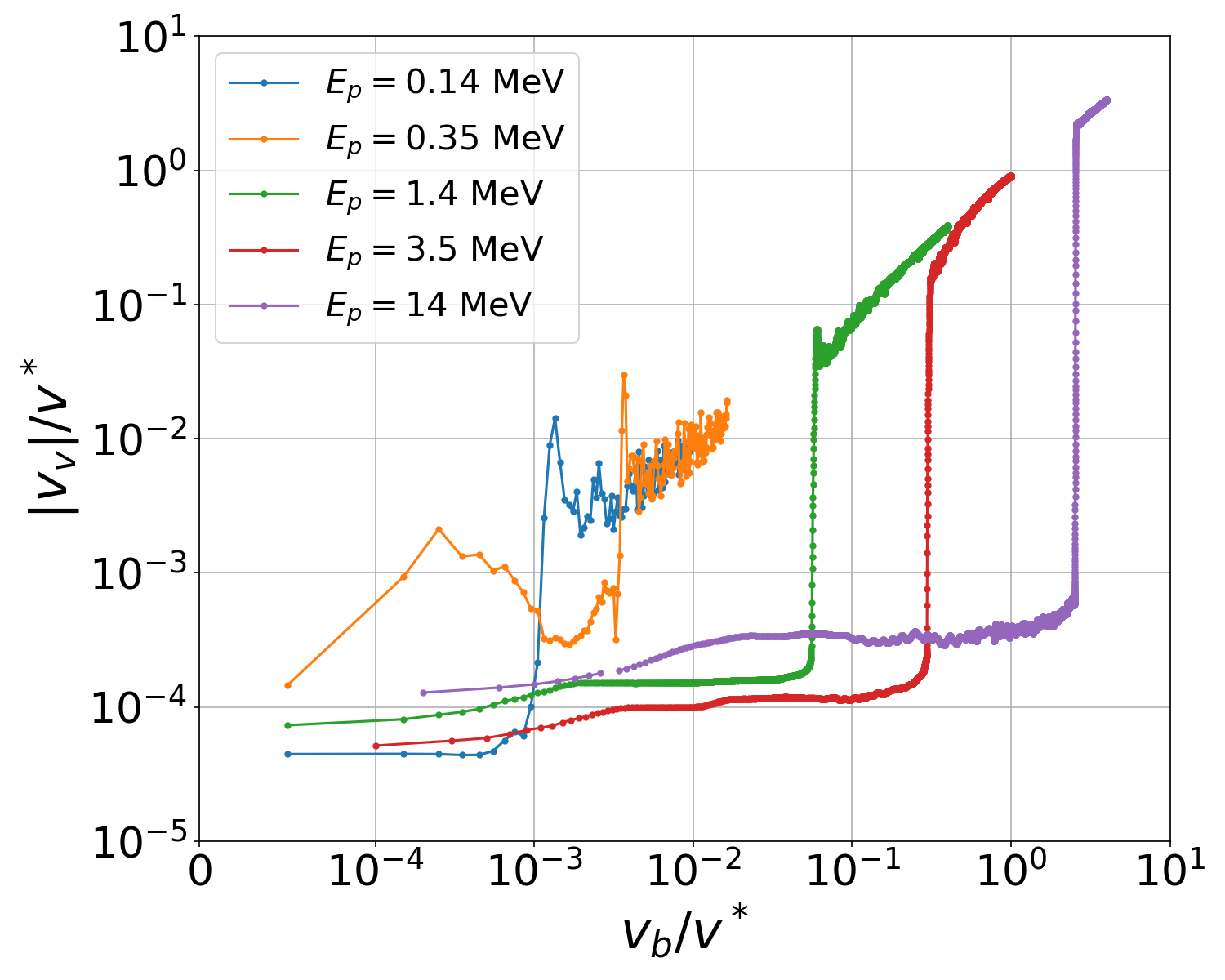}
    \caption{Vortex velocity $|v_v|$ vs. background superfluid velocity $v_b$ showing the pinning transition of the vortex for different $E_p$. The unpinning velocity depends non-linearly with the pinning energy. For low $E_p$ the tension is so strong that the vortex can't grab pinning sites and the pinning is weak; whereas at high $E_p$, the tension is negligible and the vortex passes by a greater number of nuclei, which will significantly increase the pinning strength. }
    \label{fig:E=0.4_40}
\end{figure}

The vortex pinning energy $E_p$ sets the maximum force the vortex can exert on the lattice and thus controls how much the crystal deforms. For small $E_p$, the vortex–lattice interaction is too weak to overcome Coulomb stiffness, so the lattice remains essentially ideal. As $E_p$ increases, the vortex pulls nearest-neighboring ions inward, generating more pinning sites that effectively strengthen the pinning. In the strong $E_p$ regime, this ion “clustering” deforms the lattice, enabling the vortex to attract even more nuclei into its immediate vicinity. We explore five different pinning energies, from $0.14$ MeV to $14$ MeV, with all other parameters kept the same. For reference, the critical velocity$v_c$ is $\sim 10^7$ cm/s for $E_p = 1.4$ MeV. The results are shown in \autoref{fig:E=0.4_40}. Our simulations reveal that $v_c$ increases non-linearly with $E_p$. This behavior was explored before but with fixed lattices \citep{Link_1991}. This nonlinearity arises because as $E_p$ becomes larger, the lattice and the vortex deform so that the vortex passes by a greater number of nuclei than it would in an undeformable lattice. At $E_p > 10$ MeV, the pinning is so strong that $v_c$ exceeds the typical rotational velocity of the pulsar at $\sim 10$ Hz.

\subsubsection{Repulsive interaction}
\label{subsubsec:repulsive_interaction}

For a repulse interaction $E_p<0$, a vortex pins to lattice interstices. \cite{Link_2022} showed that the interstitial pinning is weaker than the nuclear pinning by about an order of magnitude for the same value of $|E_p|$. For interstitial pinning, the vortex if farther from the nuclei on average than for nuclear pinning, so the pinning is weaker. To investigate the dynamics, we perform simulations with repulsive interaction for ten random lattice orientations. The parameters are the same as for the baseline model for attractive interaction. The simulations use lattice spacings of $b=30$ fm and $b=60$ fm. 

\begin{figure}
    \centering
    \includegraphics[width=0.98\linewidth]{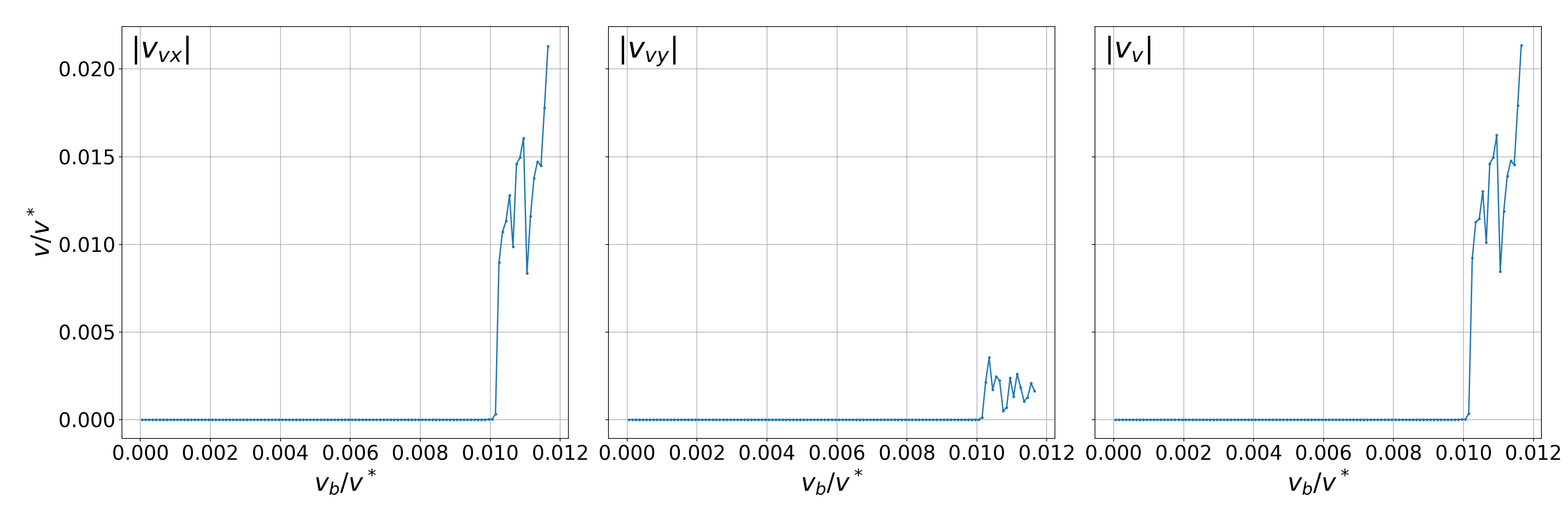}
    \includegraphics[width=0.98\linewidth]{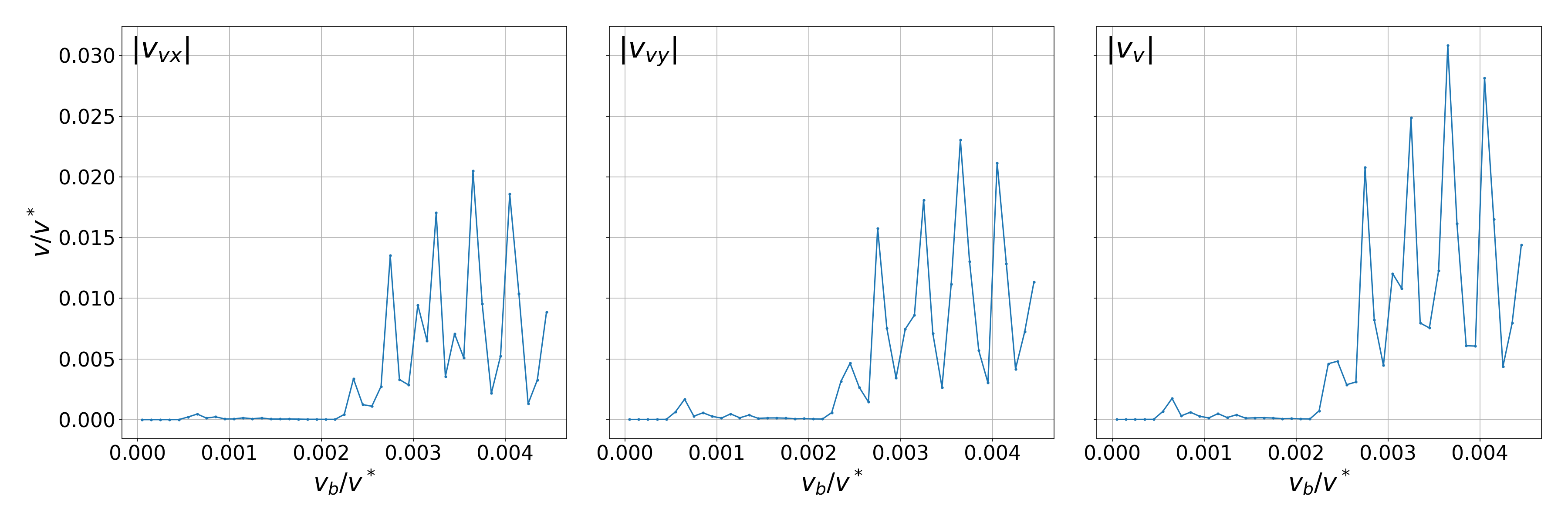}
    \caption{Vortex velocity $|v_{vx}|$, $|v_{vy}|$, and $|v_v|$ vs. background superfluid velocity $v_b$, demonstrating the unpinning of the vortex for a repulsive vortex-lattice interaction for $b=30$ fm with two different lattice orientations. The pinning is much weaker compared to that inside the lattice with an attractive vortex-lattice interaction.
    }
    \label{fig:rep}
\end{figure}

\begin{figure}
    \centering
    \includegraphics[width=0.98\linewidth]{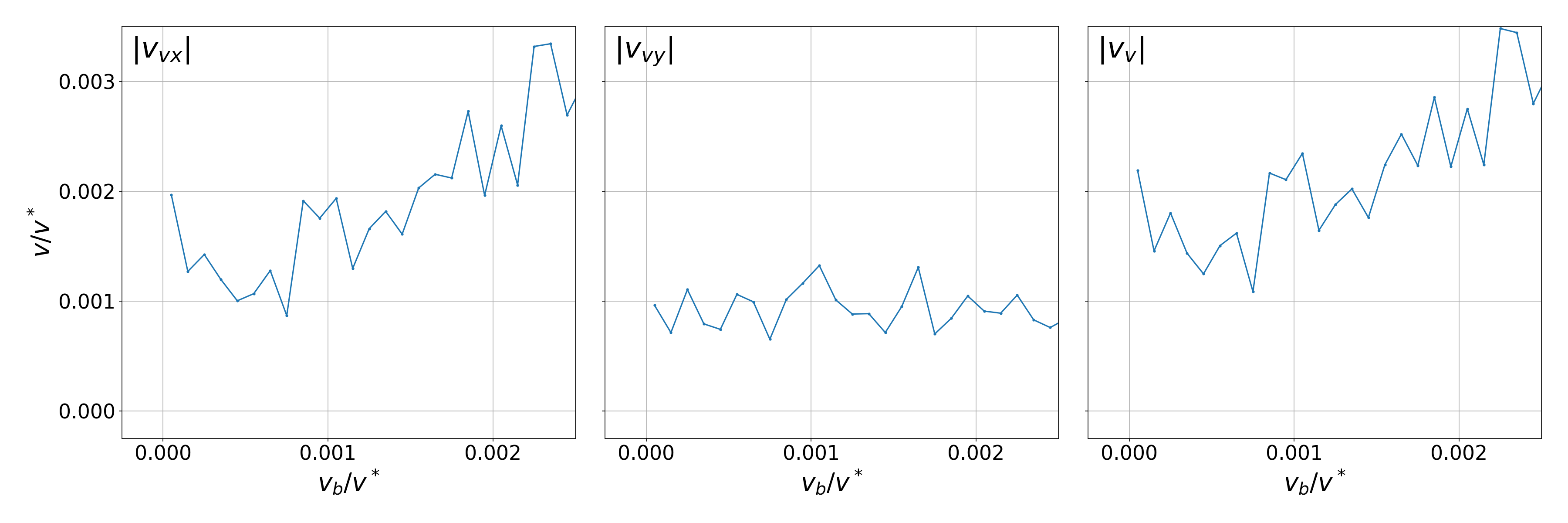}
    \caption{Similar to \autoref{fig:rep}, but with $b=60$ fm. We observe no pinning in this situation. 
    }
    \label{fig:rep_2}
\end{figure}

Two of the repulsive simulations with $b=30$ fm are illustrated in \autoref{fig:rep}. Compared to the attractive interaction, the overall pinning is much weaker for a repulsive interaction for $b=30$ fm. The unpinning superfluid velocity range is $v_c \sim 10^5 - 10^6$ cm/s, which is significantly smaller than that for the attractive pinning. Importantly, for simulations with $b=60$ fm, the vortex shows {\it no pinning at all}, as illustrated in \autoref{fig:rep_2}. In this case, the vortex is able to thread the lattice with negligible interaction with with pinning sites. 

\subsection{Vortex crossing a grain boundary}

\begin{figure}
    \centering
    \includegraphics[width=0.95\linewidth]{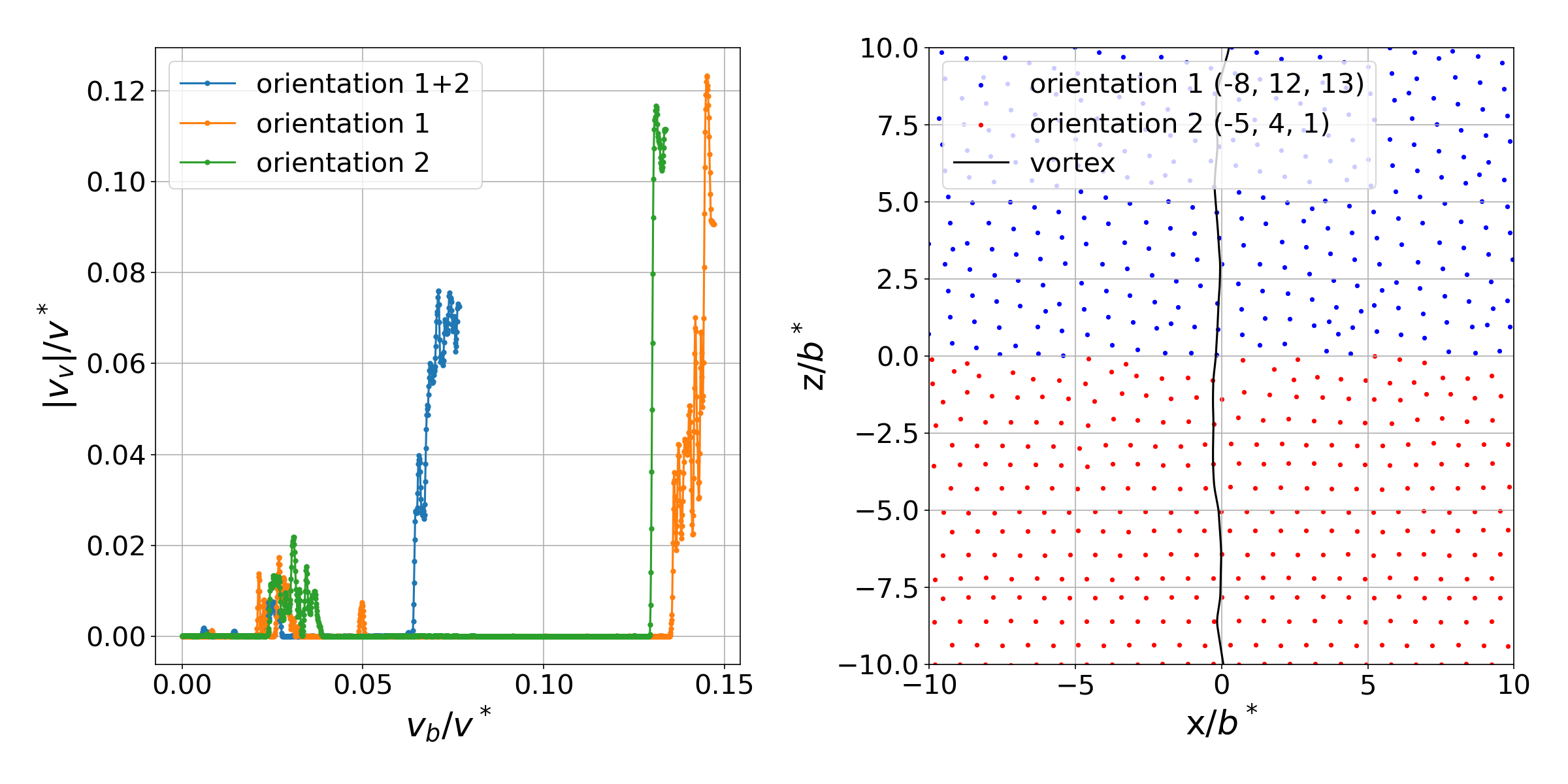}
    \includegraphics[width=0.95\linewidth]{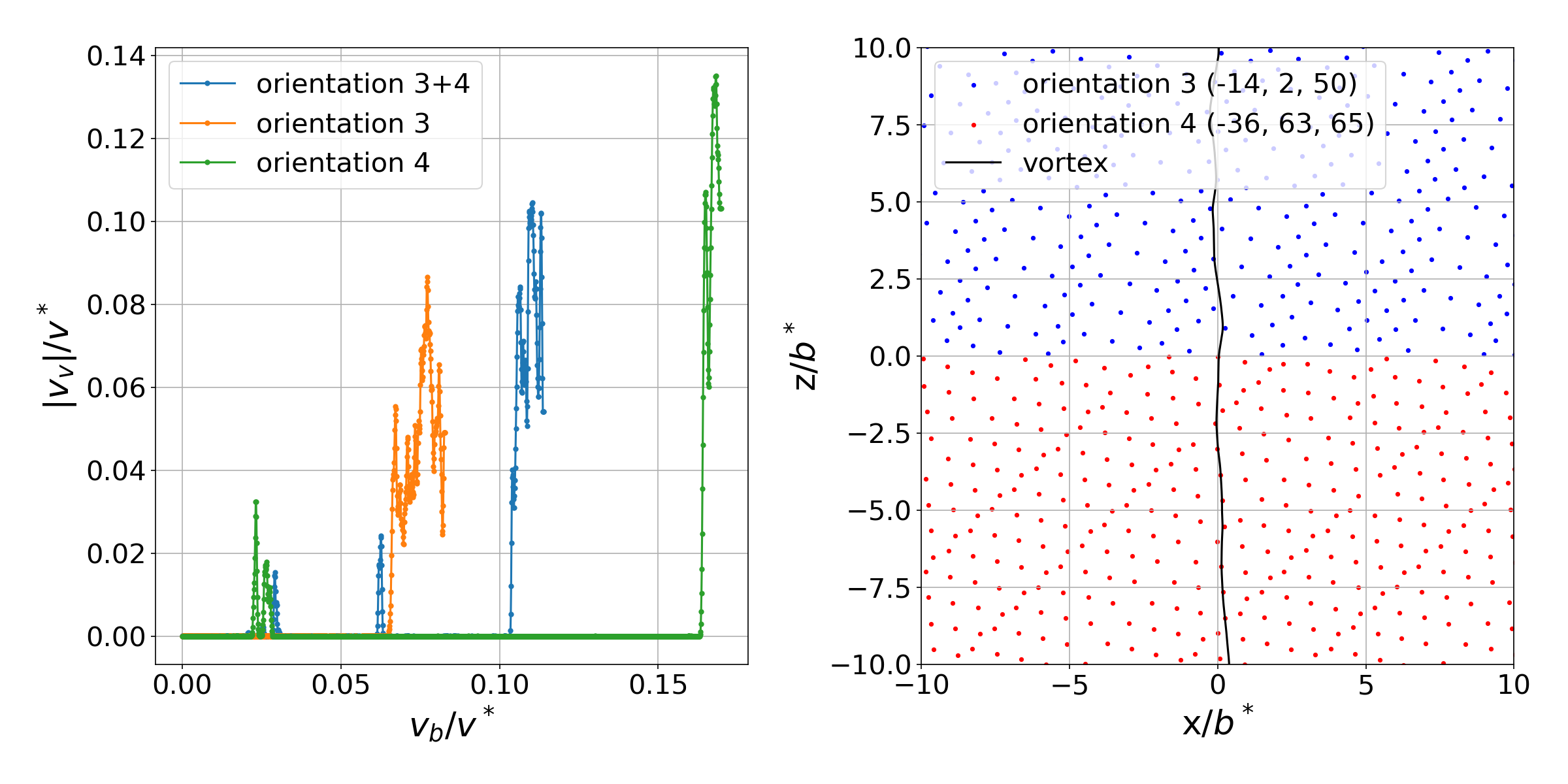}
    \caption{Left: Vortex velocity $|v_v|$ vs. background superfluid velocity $v_b$ showing unpinning of the vortex for two ajoining grains with different orientations. The grain boundary is at $z=0$. Right: zoom-in for corresponding lattice components at a slice of xz-plane ($-0.4 <y < 0.4$) and vortex element. The top plots show an example where the unpinning velocity in grains is smaller than that in both lattice orientations, whereas the bottom plots show the unpinning velocity is smaller than the that in the stronger pinning orientation. Such behavior indicates that grain boundaries can weaken the pinning and make the dynamics around the grains more complicated.}
    \label{fig:grain}
\end{figure}

The vortices in a neutron star are expected to extend through almost the entire region of S-wave pairing, which encompasses all or most of the inner crust, so the vortices traverse multiple regions (grains) of the inner crust with different lattice orientations. We now show that grain boundaries have important effects on pinning, generally weakening it. We first establish the baseline for the strength of pinning by running 4 single-lattice simulations with different lattice orientations. We then perform 6 hybrid simulation in which the top and bottom halves of the vortex are inside different grains with different lattice orientations. These grain boundaries are prepared following the methods of \cite{2025arXiv251020980C}. Two illustrative examples are shown in \autoref{fig:grain}. Grain boundaries generally have a destabilizing effects. Unpinning always initiates inside the grain with the weaker pinning or at the grain boundary. Subsequently, the unpinning propagates into the grain with stronger pinning. As a result, the effective critical velocity of the hybrid system falls below that of the strong-pinning grain, sometimes below the mean of the two grains, and can even be lower than that of the weakly-pinning grain. Such results indicate that on a macroscopic scale, it is incorrect to average the critical velocity over the lattice orientations, as was done in \cite{Link_2022}. This is particularly important for the case of pinning, in which case there can be lattice orientations where the pinning is so weak that it is not measurable in numerical experiments.We  perform grain boundary simulations for repulsive interactions and observe how the vortex unpins from the region with weaker pinning; the unpinning wave quickly propagates into the region with stronger pinning. 

\subsection{Vortex in a lattice undergoing shear}

\begin{figure}
    \centering
    \includegraphics[width=0.98\linewidth]{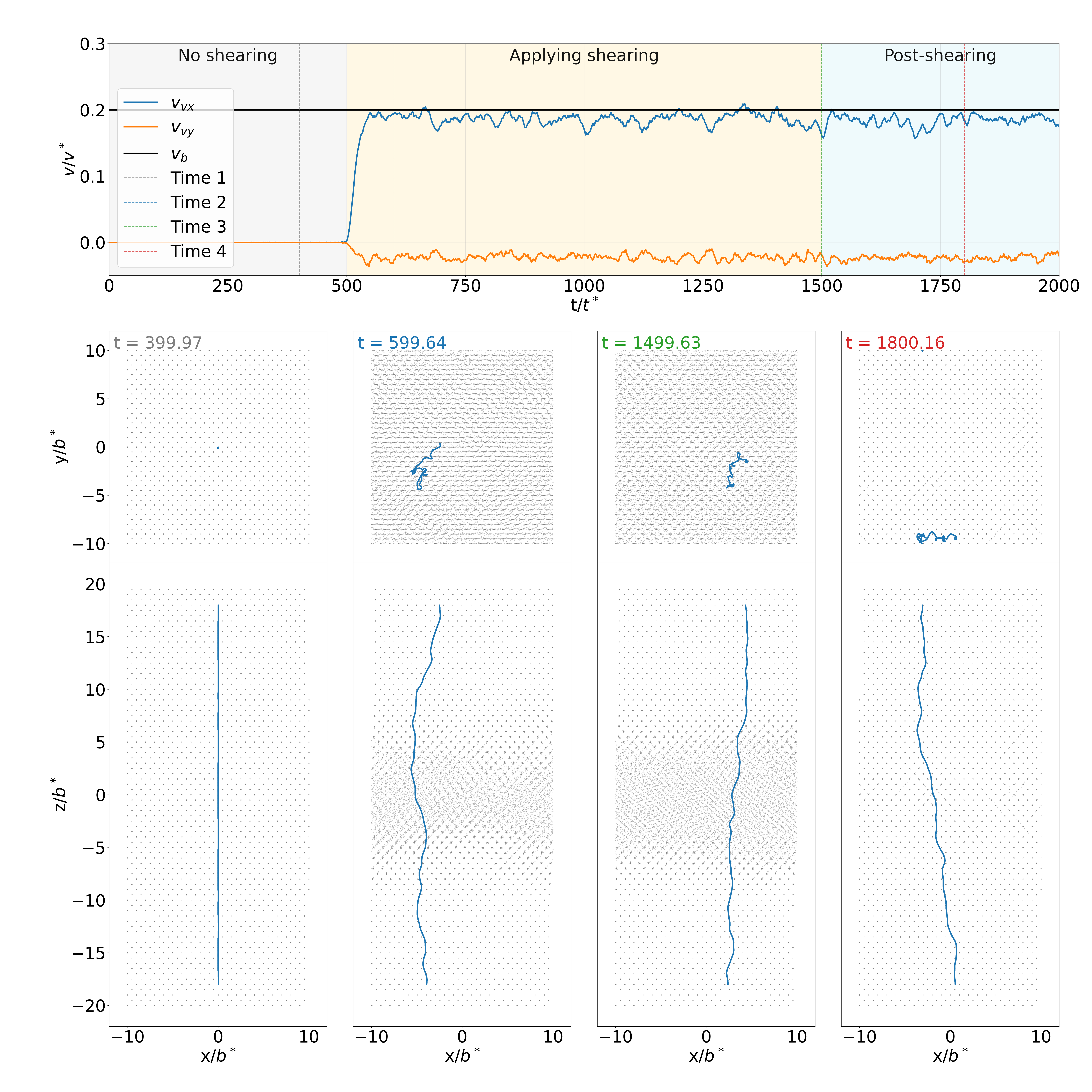}
    \caption{Vortex motion in attractive interaction with lattice shearing. Top: vortex velocity $v_v$ at x and y direction vs. time. Bottom: four snapshots in selected time frame viewing from $xy$- and $xz$-plane. The background superfluid velocity $v_b$ is indicated as black solid line in the top plot. The lattice is initially at rest and $v_b/v^* = 0.2$, which is above the repinning velocity $v_r$ and below the unpinning velocity $v_c$. The vortex is pinned initially. From $t=500t^*$, the lattice at $z>10$ starts moving to $+x$ direction, while the lattice at $z<10$ moves to $-x$ direction. The vortex becomes unpinned because the shearing excites the Kelvin wave and moves the pinning sites. The shearing is stopped at $t=1500t^*$ and the lattice is able to re-crystallize. However, the vortex is not repinned because it's already above the repinning threshold. Therefore, shearing can cause the vortex unpinning as long as $v_b > v_r$. }
    \label{fig:shear_1}
\end{figure}

\begin{figure}
    \centering
    \includegraphics[width=0.98\linewidth]{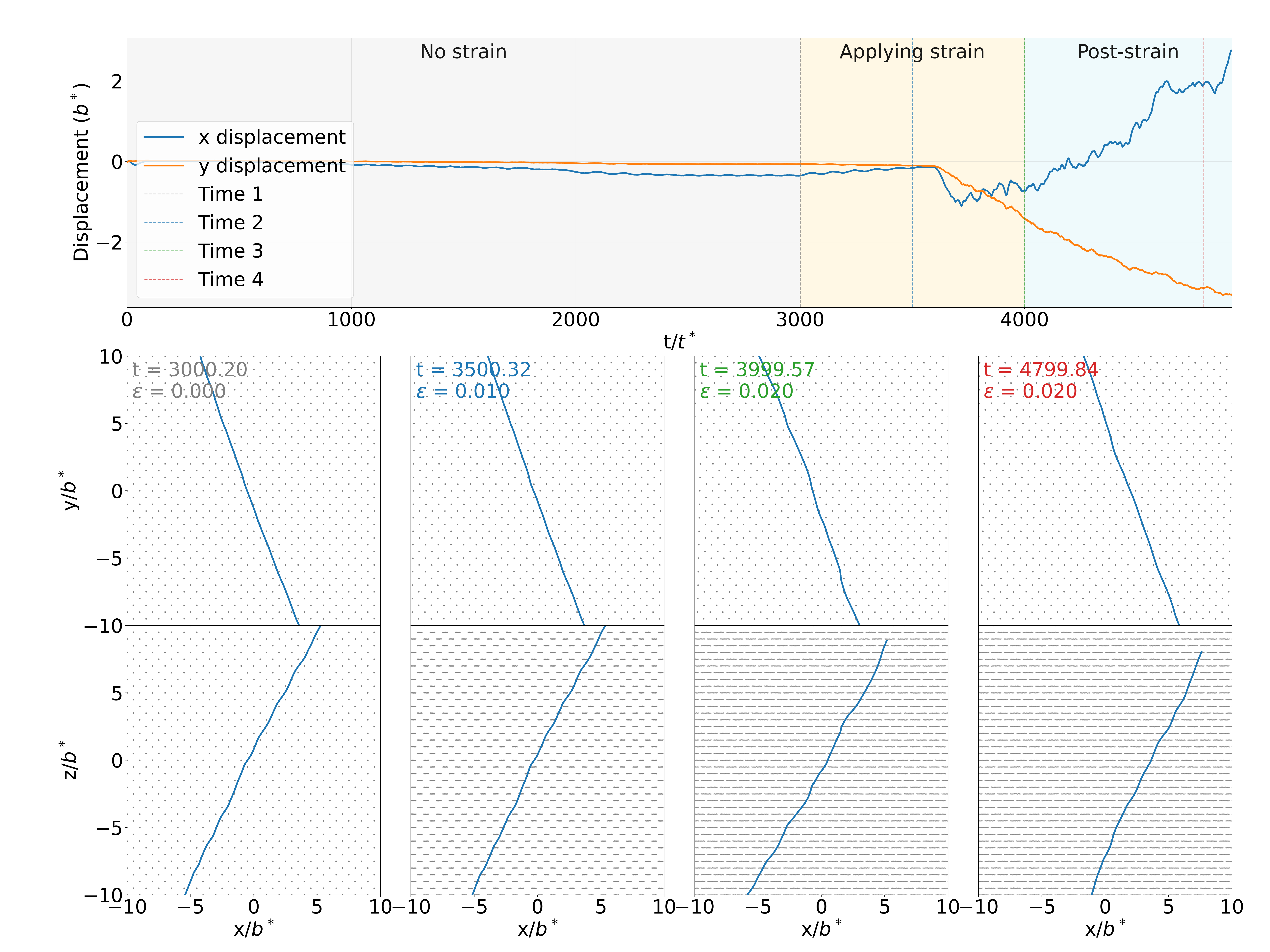}
    \caption{Vortex motion for a repulsive vortex-nucleus interaction for a lattice under small strain. The deformation starts at $t=3000t^*$ and ends at $t=4000t^*$, with the maximum strain $\epsilon = 0.02$. The background superfluid velocity $v_b/v^*=0.01$, which is smaller than the critical unpinning velocity $v_c$. Top: vortex displacement at x and y direction vs. time. Bottom: four zoom-in snapshots in selected time frame viewing from $xy$- and $xz$-plane. The vortex keeps moving after the lattice deformation, indicating that small strain cause unpinning.}
    \label{fig:shear_2}
\end{figure}

Magnetic or spin-down stresses in an evolving neutron star could shear the crust and cause it to fail, perhaps triggering the release of a large number of vortices.
In this section, we model a stress-induced shear in a lattice and demonstrate the unpinning of vortices that are crossing the strongly sheared region. We find that the unpinning takes place when the background superfluid velocity exceeds the critical velocity for {\it re-pinning} of the vortices, but is still smaller than the critical velocity for {\it unpinning} of the otherwise unperturbed vortices. In other words, we find that a strong lattice shear is able to change the vortex state and ``kick" the vortex from the lower to the upper branch of the hysteresis loop (see \autoref{fig:baseline})
\footnote{While the dynamics of the strongly sheared and structurally failing lattice is an interesting topic in its own right, we do not discuss it here in any detail.}.

The simulation is performed as follows: we start with the vortex aligned with the z-axis of the lattice, and introduce a shear field $f(z,t)\hat{x}$ to the lattice, where $f(z,t)$ is the lattice displacement; $f(z,t)<0$ for $z<0$, and $f(z,t)>0$ for $z>0$. We show an example simulation in \autoref{fig:shear_1}. The vortex is initially pinned since the background superfluid velocity $v_b$ is smaller than the unpinning velocity $v_c$. At $t=500t^*$, we initiate the shearing motion in $\pm x$ directions. The vortex quickly unpins. At $t=1500t^*$, we stop the shearing. The vortex remains unpinned and continues moving. 

In addition, we simulate a smaller shearing event for a repulsive pinning. We apply a small strain with maximum $\epsilon = |df/dz|_{\mathrm{max}} = 0.02$, and the vortex becomes unpinned (see \autoref{fig:shear_2}). Interstitial pinning is more fragile against the lattice shearing.

Unpinning can be initiated even when the imposed background velocity $v_b$ remains below the critical velocity $v_c$. The lattice deformation excites Kelvin waves, which propagate along the vortex. Our simulations demonstrate the importance of lattice dynamics for vortex unpinning in neutron star crusts and suggests that starquakes, if they occur, could precipitate glitch events without requiring the superfluid to spin up to its critical unpinning velocity. This provides a possible mechanism for simultaneously  unpinning all of the vortices that initially cross the slip plane of the quake. It is an open question whether magnetic stresses \citep{2025ApJ...979..144B} or spin-down stresses \citep{Ruderman_1991} inside the crust of a pulsar can cause breaking of the crust at some location.

\subsection{Double Vortex Interaction}

\begin{figure}
    \centering
    \includegraphics[width=0.85\linewidth]{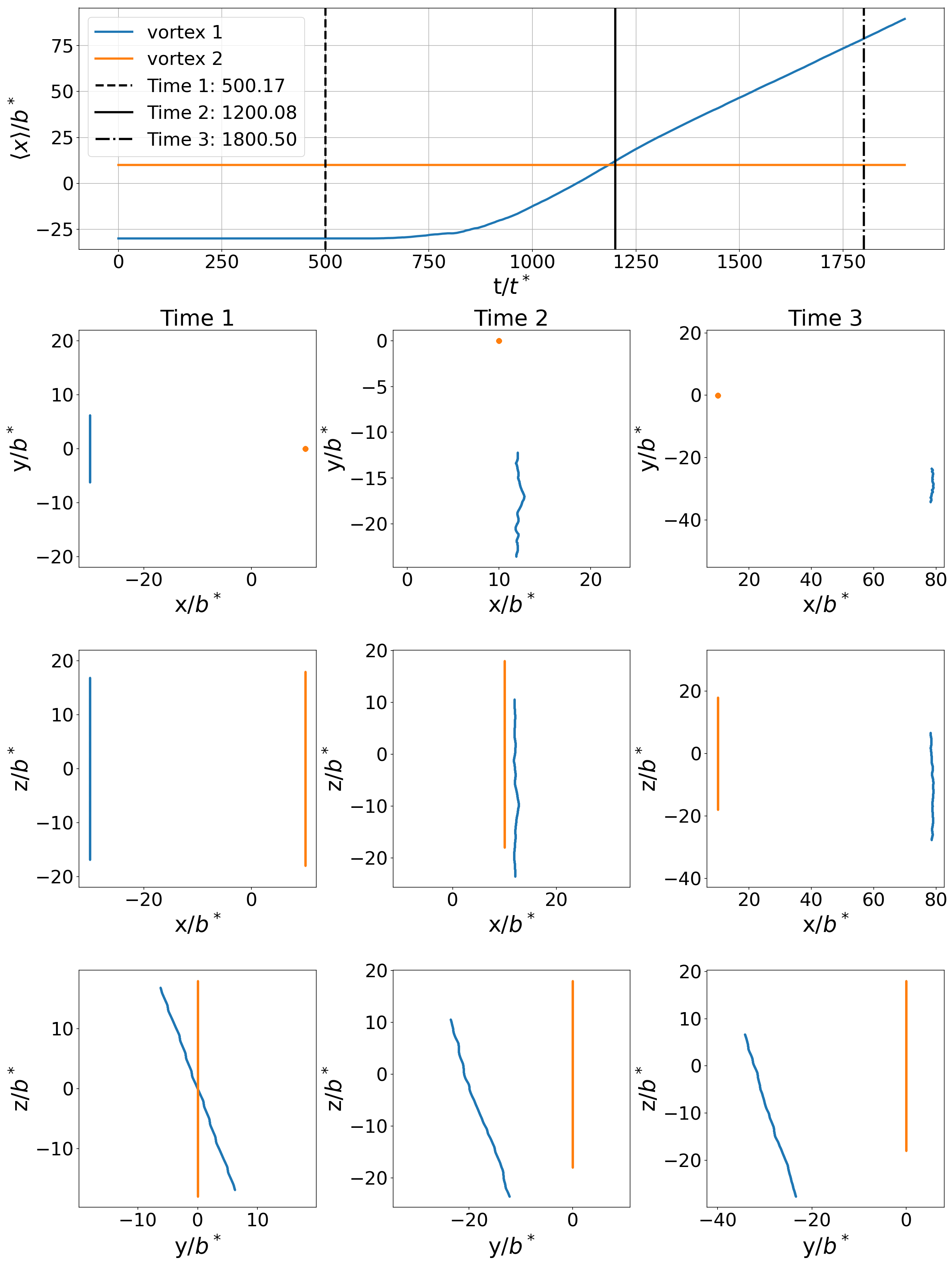}
    \caption{Double vortex simulation showing one vortex passing a second vortex without unpinning it (nuclei are not shown). Top: Vortex average $x$ position $\langle x \rangle$ vs. time for vortex 1 and 2, showing the mobile vortex coming close to and then bypassing the pinned vortex. Bottom: three snapshots from $xy$-, $xz$-, and $yz$-planes for the vortices at selected time steps. Each column shows the same timestep. }
    \label{fig:double_1}
\end{figure}

\begin{figure}
    \centering
    \includegraphics[width=0.85\linewidth]{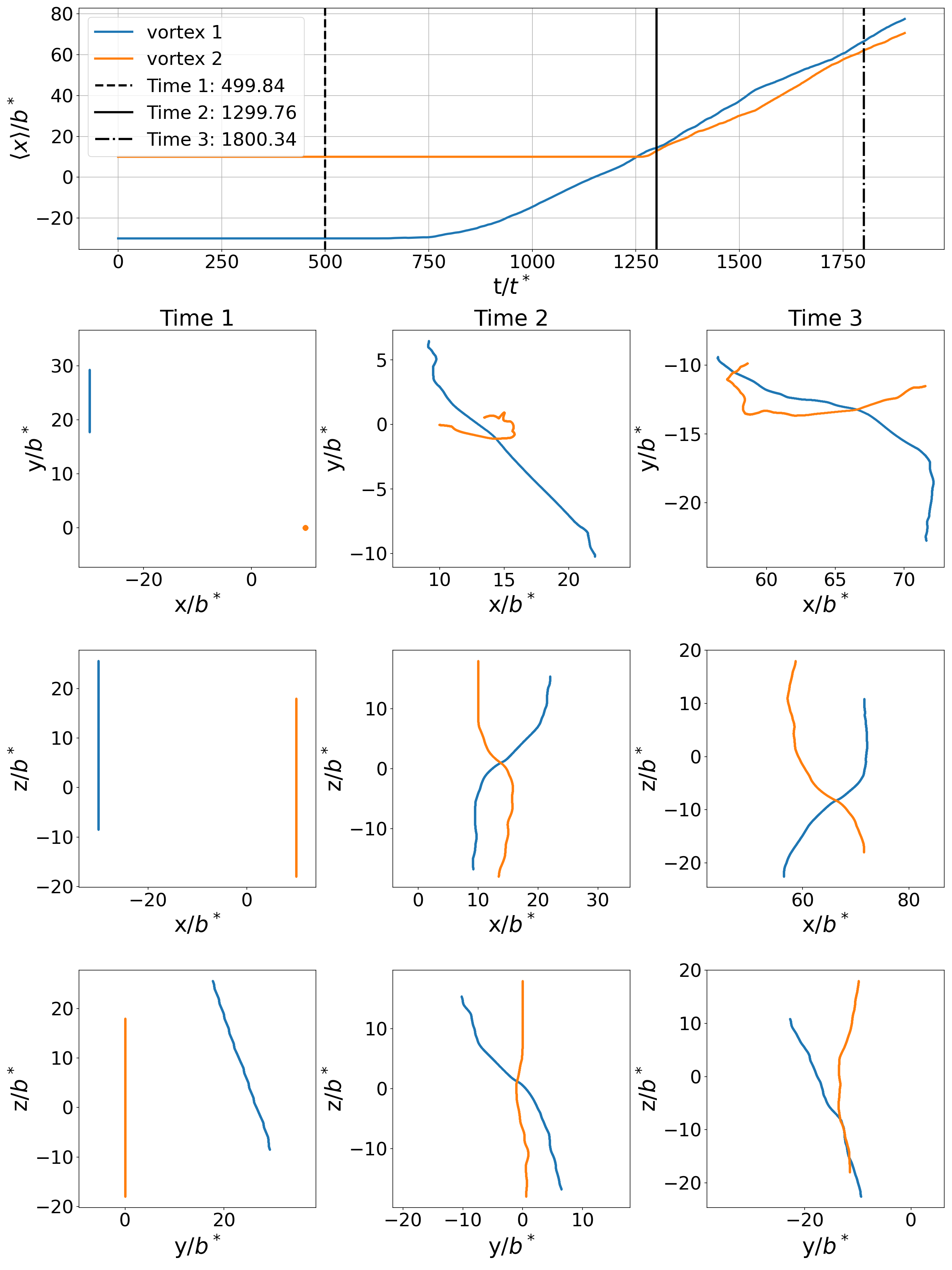}
    \caption{Similar to \autoref{fig:double_1}, but with the mobile vortex unpinning the initially stationary pinned vortex. The first column shows the vortex position before the unpinning. The second column shows vortex 1 causing vortex 2 to unpin. The third column shows the dynamics after the unpinning.}
    \label{fig:double_2}
\end{figure}

In a spinning-down neutron star the pinned vortex array could be driven into complex configurations \citep{Link_2022, Levin_2023}. Migrating unpinned vortices are likely to come near pinned vortices. We now explore whether a moving vortex can unpin a neighboring vortex when they nearly intersect. 

In two-vortex simulations, the superfluid velocity at the first vortex $\bm{v_{s1}}$ has an extra term contributed by the second vortex given by the Biot-Savart law,

\begin{equation}
\label{eqn:back_vel}
    \bm{v_{s1}}(z) = \bm{v_b} + \bm{v_2}(z) = \bm{v_b} + \frac{\kappa}{4\pi}\int_{\bm{u_2}} \frac{\bm{u_2}-\bm{u}}{|\bm{u_2}-\bm{u}|^3} \times d\bm{u}_2,
\end{equation}

\noindent where the first term $\bm{v_b}$ is the background superfluid velocity and the second term $\bm{v_2}$ is the induced velocity field generated by another vortex; here $\bm{u}_2$ denotes a point on the second vortex. Similarly, $\bm{v_{s2}}(z) = \bm{v_b} + \bm{v_1}(z)$. To get an idea of the magnitude of the $\bm{v_2}$, we compute the velocity field generated by an infinite straight vortex at distance $r$ from the vortex core, and equate it to the background flow $v_b$::
\begin{equation}
\label{eqn:induced_velocity}
    |\bm{v_2}(r)| = \frac{\kappa}{2\pi r} = 10^8 \ \left(\frac{30 \ \mathrm{fm}}{r}\right) \  \mathrm{cm/s}.
\end{equation}
We compare $v_2$ with the characteristic background unpinning velocity $v_c\sim 10^7\mathrm{cm/s}$ to estimate when $v_2$ is significant. For $v_2 \sim v_c$, the distance $r \sim 10^2$ fm, which is orders of magnitude smaller than the average spacing between vortices of $\sim 10^{-2}$ cm. Therefore the vortices will affect each other only if they nearly intersect; this justifies the focus of this section on the dynamics of two nearly-intersecting vortices. 

Even for two vortices, the computation is expensive, since the complexity of the computation is $O(n^2)$ in each timestep, where $n$ is the number of the vortex elements. To minimize the computational cost, we adopt the tree method introduced in \cite{Baggaley_2012} to handle the vortex-vortex interaction. This method decreases the computational complexity to $O(n \log{n})$ by approximating the remote vortex elements as a point sources while still fully resolving the closer vortex elements for better precision. Further details can be found in \autoref{Appendix:tree}. To minimize computation time, we use a static lattice and introduce a linear drag force $\eta \partial \bm{u}/\partial t$ on the vortex with $\gamma=\eta/\rho_s\kappa=0.1$. 

Initially both vortices are perpendicular to the superfluid flow, but are not parallel to each other. The initial distance between two vortices in x-direction is $\approx 40b^*$. After relaxation, we impose a background superfluid velocity $v_b$ in the $+x$ direction that is larger than the critical velocity for vortex 1, but smaller than that for vortex 2. Therefore, vortex 1 is unpinned while vortex 2 initially remains pinned. We intentionally select the initial location of vortex 1 so that it approaches vortex 2 after unpinning. The induced velocity at vortex 2 generated by vortex 1 diverges when the distance between two vortices $r$ goes to zero. To avoid this problem, we set a vortex core radius equal to the equal to the typical neutron coherence length $r_{\mathrm{core}}=5$ fm in the inner crust \citep{Blasio_1997} and set a minimum distance $r_{\mathrm{min}} = 2r_{\mathrm{core}}$. Our simulations reveal two possible outcomes when the two vortices interact:

(1) {\it Vortex 1 passes vortex 2 without unpinning it}, as in \autoref{fig:double_1}. This is because in our finite‐length simulations, the moving vortex may avoid the neighboring vortex and continue past it. This will happen if the unpinning velocity of vortex 2 is larger than $|\bm{v_{s2}}|$. Such avoidance seems unlikely for macroscopically long vortices.

(2) {\it Unpinning of vortex 2}, as in \autoref{fig:double_2}. This will happen if the unpinning velocity of vortex 2 is smaller than  $|\bm{v_{s2}}|$ at some location on vortex 2. The unpinning is initiated near the the point of closest approach between the two vortices, and then quickly propagates along the length of the vortex 2. It's also possible that two vortices reconnect; however we cannot resolve this using our approach. It has been shown that reconnection launches strong Kelvin waves on both of the reconnecting vortices \citep{Schwarz_1978}. If the background superfluid velocity is larger than the repinning velocity of vortex 2, then the whole of vortex 2 will be unpinned.

\section{Discussion}
\label{sec:discussion}

Vortex drag plays a crucial role in vortex pinning and unpinning. In previous work (e.g., \cite{Link_2022}), a drag force linear in velocity was introduced in the vortex equation of motion to model drag against the nuclear lattice. The drag coefficient was unknown and kept as a free parameter. Through simulations of vortex motion coupled to the dynamic lattice, we have determined the drag force from first principles. As a vortex moves, Kelvin waves are excited, which are then dissipated by coupling to lattice vibrations, producing drag on the vortex. We have confirmed the hysteresis behavior observed in \cite{Link_2022}, but we have found that it is strongly modified by thermal effects, which weaken pinning for crust temperatures above $10^8$K. We have also found that the grain structure considerably weakens pinning of long vortices that traverse the grains, and the pinning is especially weakened for a repulsive interaction.

Our simulations also give insight on how quantum vorticity may become unpinned in macroscopically large regions of the crust. First, we have shown that strongly sheared planar regions (such as expected during 
a crustal failure) trigger the unpinning of the vortices that cross it. Subsequently, the pattern of unpinnings  propagates along the vortices, thus forming highly elongated regions of unpinned vorticity. Secondly, we have found that while the vortices are on average far away from each other
(compared to their core size), they can interact strongly when they nearly intersect. An unpinned vortex passing near a pinned vortex often causes the latter to unpin, with the subsequent pattern of unpinnings propagating along the latter vortex. Since vortices are orders of magnitude longer than the distance between them, their intersections are likely unavoidable and thus vortex avalanches might develop.

Extended global simulations are needed to capture the collective behavior of vortex arrays over longer timescales required for the pattern of unpinnings to propagate over macroscopic distances. Construction of such models is the subject of our future work.

\begin{acknowledgments}

This work was supported by a grant from the Simons Foundation (MP-SCMPS-00001470) to MC and YL, and by YL's Simons Investigator Grant $827103$. 
Financial support for this publication also comes from Cottrell Scholar Award \#CS-CSA-2023-139 sponsored by Research Corporation for Science Advancement,  
and from the National Science Foundation under Grant No. NSF PHY-1748958. MC thanks the KITP for hospitality and MC acknowledges support as a KITP Scholar. 
\end{acknowledgments}

\appendix

\section{solving vortex equation of motion}
\label{Appendix:solver}

The solution closely follows \cite{Link_2022} and is summarized here. We normalize \autoref{eqn:eom} with length unit $b^* = 30$ fm and time unit $t^* = \rho_s \kappa b^{*2}/T_v =1.86 \times 10^{-20}$ s. Define $\psi \equiv u_x + iu_y$ and $\phi \equiv f_{Lx} + if_{Ly}$,
\begin{equation}
\label{eqn:reduced}
    \frac{\partial^2\psi}{\partial z^2} + i \left(\frac{\partial \psi}{\partial t} - v_s\right) + \frac{b\phi}{T_v} = 0,
\end{equation}

We then expand $\psi$ in Fourier series with free‐end conditions $\partial_z\psi|_{z=0,L}=0$.,

\begin{equation}
\label{eqn:fourier}
    \psi = \sum_0^\infty a_n(t) \cos{k_n z}, \ k_n = n\pi/L
\end{equation}

Substitute \autoref{eqn:fourier} into \autoref{eqn:reduced} and make use of Fourier series orthogonality,

\begin{equation}
\label{eqn:a_n_1}
    \dot{a}_n + ik_n^2 a_n = \bar{v}_{n} + i\bar{\phi}_n,
\end{equation}

where 

\begin{align}
\label{eqn:FV}
    \bar{\phi}_n(T) & = \frac{2b}{LT_v}\int_0^L \phi(z, t) \cos{(k_n z)} dz \\
    \bar{v}_n(T) & = \frac{2}{L}\int_0^L v_s(z, t) \cos{(k_n z)} dz \\
\end{align} for $n>0$, and
\begin{align}
\label{eqn:FV_2}
    \bar{\phi}_0(T) & = \frac{b}{LT_v}\int_0^L \phi(z, t) dz \\
    \bar{v}_0(T) & = \frac{1}{L}\int_0^L v_s(z, t) dz. \\
\end{align}

To handle the non-linear term on the RHS of \autoref{eqn:a_n_1}, we apply an exponential integrator

\begin{equation}
\label{eqn:a_n}
    \frac{dQ(t)}{dt} \equiv \frac{d}{dt}[a(t)e^{ik_n^2t}] = (\bar{v}_{n} + i\bar{\phi}_n)e^{ik_n^2t}.
\end{equation}

Spatially, we discretize $z$ with $\Delta z=0.01$ ($N_z=L/\Delta z$ points; for $L=36b^*$, $N_z=3600$) and retain $N_n=4L$ Fourier modes to ensure spectral accuracy. The advantage of the exponential integrator is that it will give an exact solution when $\bar{\phi}_n$ of \autoref{eqn:a_n} is zero. We then apply a fifth-order Runge-Kutta Dormand–Prince method (RKDP) with adaptive step size to control error. 

A test can be performed by calculating the unpinning velocity of a straight vortex without any misalignment. In that case, an analytical unpinning velocity can be obtained by integrating the interaction force over the vortex,

\begin{align}
    f_{L}(r) & = \frac{2\sqrt{\pi}E_ir}{b\sigma}\exp{\left(-\frac{r^2}{\sigma^2}\right)} \\
    f_{L\mathrm{, max}} & =  \frac{E_i}{b}\sqrt{\frac{2\pi}{e}}\\
    v_c/v^* & = \frac{bf_{L\mathrm{, max}}}{T_v}
\end{align}

For $b=b^*$, our calculation shows $v_c/v^* = 0.34$; in the simulation, we get the same result. 

\section{Computations of vortex-vortex interaction}
\label{Appendix:tree}

The vortex-vortex interaction induced velocity $v_2$ is given in \autoref{eqn:back_vel}. Specifically, for a straight vortex element $u(z_{j+1}-z_j)$, we can compute its $\bm{v_2}$ to $u_i$ as \citep{Baggaley_2012}
\begin{equation}
\label{eqn:vd}
    \bm{v_{2,ij}} = \frac{\kappa}{2\pi (4AC-B^2)} \left[\frac{2C+B}{\sqrt{A+B+C}} - \frac{B}{\sqrt{A}}\right] \bm{p} \times \bm{q},
\end{equation}

\noindent where $\bm{p} = \bm{u_j}-\bm{u_i}$, $\bm{p} = \bm{u_{j+1}}-\bm{u_j}$, $A=|\bm{p}|^2$, $B=\bm{p} \cdot \bm{q}$, and $C=|\bm{q}|^2$. In principle, we can repeat this calculation for all vortex elements. However, such calculation require $O(n^2)$ complexity, which can be expensive for large $n$.

To reduce the complexity, we apply tree method introduced in \cite{Baggaley_2012}, which we briefly summarize below. Given a discrete vortex $\bm{u}(\Delta z)$, we create a tree by continuously dividing the vortex by half and generate a node correspondingly, until each node only contain one vortex element. We start from the largest node and calculate the total circulation $\bm{u_t}$ and the center of circulation $\bm{\bar{u}}$ of the node
\begin{equation}
    \bm{u_t} = \sum_{j=1}^N (\bm{u_{j+1}} - \bm{u_j}),
\end{equation}
\begin{equation}
    \bm{\bar{u}} = \frac{1}{N}\sum_{j=1}^N \bm{u_j}.
\end{equation}
Now we want to compute the circulation at $\bm{u_i}$. We first compute the opening angle $\theta \equiv L/d$, where $L$ is the length of the vortex and $d$ is the distance between the $\bm{\bar{u}}$ and $\bm{u_i}$. We define a critical opening angle $\theta_{max}$. If $\theta<\theta_{max}$, we accept the node and compute the circulation using \autoref{eqn:vd}, using $\bm{p} = \bm{\bar{u}}$ and $\bm{q} = \bm{u_t}$. If $\theta>\theta_{max}$, we open the node and repeat the test for each child node, until only one element left in child node or $\theta<\theta_{max}$. Such calculation has complexity $O(n\log{n})$, depending on the choice of $\theta_{max}$. We apply $\theta_{max} = 1.0$ in our simulation. 


\bibliography{sample631}{}
\bibliographystyle{aasjournal}



\end{document}